\documentstyle[epsfig,12pt]{article}
\begin{document}

\def\emline#1#2#3#4#5#6{%
       \put(#1,#2){\special{em:moveto}}%
       \put(#4,#5){\special{em:lineto}}}
\def\newpic#1{}

\begin{titlepage}

\hfill December 11, 2010

\vspace*{2cm}
\begin{center}
\Large \bf Threshold corrections to the MSSM finite-temperature
            Higgs potential
\end{center}

\vskip 2cm

\begin{center}
\large
        M.~Dolgopolov$^\sharp$,
        M.~Dubinin$^*$,
        E.~Rykova$^\sharp$
\end{center}

\begin{center}
 { \it $^\sharp$ Samara State University,
                  443011 Samara, Russia}   \\
 { \it $^*$ Skobeltsyn Institute of Nuclear Physics, Moscow State
                  University, 119991 Moscow, Russia}
\end{center}

\begin{quotation}

\vskip 2cm
\begin{center}
{\sf Abstract}
\end{center}

\vskip 0.5cm

In the minimal supersymmetric standard model (MSSM) 
the one-loop finite-temperature corrections from 
the squarks-Higgs bosons sector are calculated, the effective 
two-Higgs-doublet potential is reconstructed and possibilities of the electroweak 
phase transition in full MSSM ($m_{H^\pm}$, ${\tt tg}\beta$, $A_{t,b}$, 
$\mu$, $m_Q$, $m_U$, $m_D$) parameter space are studied. At large values of $A_{t,b}$
and $\mu$ of around 1 TeV, favored indirectly by LEP2 and Tevatron data, 
the threshold finite-temperature corrections from triangle and box diagrams 
with intermediate third generation squarks are very substantial. 
Four types of bifurcation sets are defined for the two-Higgs-doublet potential. 
High sensitivity of the low-temperature evolution to the effective 
two-doublet and the MSSM squark sector parameters is observed, but rather 
extensive regions of the full MSSM parameter space allow the first-order 
electroweak phase transition respecting the phenomenological constraints 
at zero temperature. As a rule, these regions of the MSSM parameter space are in
line with the case of a light stop quark. 
\end{quotation}

\vskip 1.5cm

PACS:


\end{titlepage}
\section{Introduction}
\vskip 0.2cm

\noindent
The absence of antimatter in the Universe (the baryon asymmetry), a small
ratio of the observed number of baryons to the observed number of photons
$n_B/n_{\gamma} \sim$ 6 $\times$ 10$^{-10}$ and the absence of light 
($m_H \sim$100 GeV) CP-even Higgs boson signal at LEP2 and 
Tevatron energies lay a specific 
claim to models of particle physics. The baryon asymmetry and an 
extremely small $n_B/n_{\gamma}$ could be understood on the basis of 
Sakharov conditions, which are respected at the electroweak phase 
transition, expected to take place at the temperature of the 
order of 10$^2$ GeV \cite{ewtrans_gen}. Generation of nonzero vacuum 
expectation value $v$ 
of the scalar field breaks the electoroweak symmetry $SU(2)_f \times 
U(1)_{Y}$ to the electromagnetic symmetry $U(1)_{em}$. It 
is well-known \cite{kirzhnits} that in the simple isoscalar model with the 
standard-like Higgs potential
$U(\varphi)=-\frac{1}{2}\mu^2 \varphi^2 + \frac{1}{4}\lambda \varphi^4$,
describing a thermodynamically equilibrium system of the scalar particles 
at the temperature $T$, the equation for the vacuum expectation value
$v(T)$ has two solutions: $v(0)=0$ and 
$v^2(T)=\mu^2/\lambda-T^2/4$, demonstrating the second order phase 
transition at the critical temperature $T_c=2\mu/\sqrt{\lambda}=2v(0)$, 
see Fig.1a.
The thermal Higgs boson mass $m^2_h= -\mu^2+\lambda T^2/4$ vanishes at
the critical temperature $T_c$ thus restoring the spontaneously broken 
symmetry. 

\unitlength 1.00cm
\begin{figure}[h!]
\linethickness{0.4pt}
\begin{center}
\begin{picture}(10,5.2)
\put(-2.5,-5.5){\epsfxsize=14cm \epsfysize=12cm \leavevmode 
\epsfbox{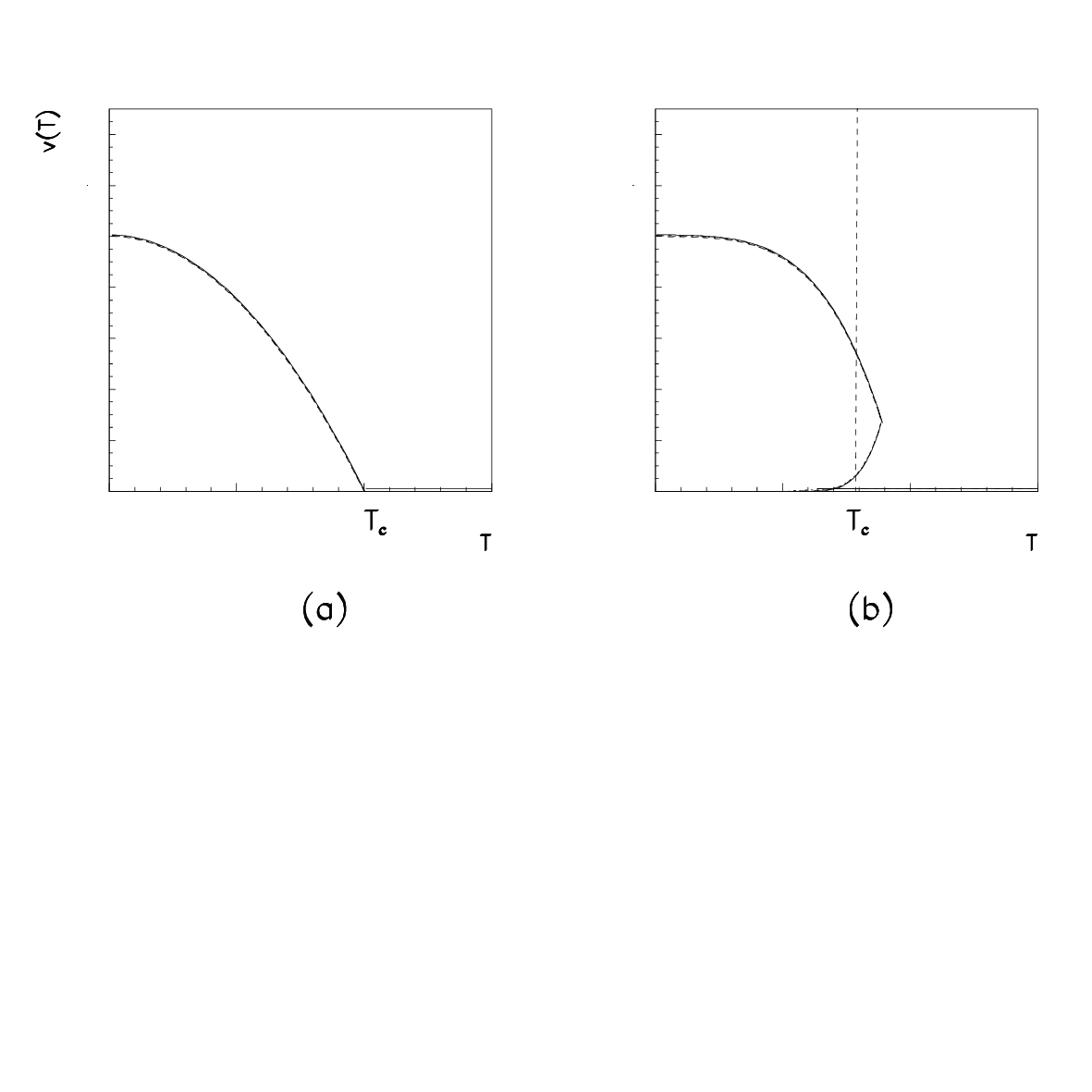}}
\end{picture}
\end{center}
\caption{\small
Contours on the $(v,T)$ plane for (a) the second order phase 
transition (fracture point) and (b) the first order phase transition at 
the critical temperature $T_c$ (dashed vertical line).} 
\end{figure} 

However, in the cosmological evolution the stages with thermodynamically
non equilibrium plasma and the first order phase transitions (see 
a typical $v(T)$ contour in Fig.1b) are very important, so such simple 
picture in 
combination with the standard model CP-violation by means of the CKM mixing
matrix turns out to be not sufficient to justify the observed ratio of baryon
number to entropy. 
The situation becomes better in the minimal
supersymmetric model (MSSM) where sparticles, extended two-doublet Higgs 
sector with the two background fields and 
nonstandard sources of CP-violation provide a number of new possibilities. 
In a number of approaches \cite{approaches} the electroweak phase
transition is defined by evolution of the finite temperature effective
Higgs potential involving the cubic term in the background scalar fields 
$v_1, v_2$. The
larger this term is, the stronger pronounced turns out to be the first
order phase transition, which is essential for consistency with the
Higgs boson mass beyond the LEP2 exclusion $m_H <$115 GeV. Enhancement of
the cubic term in the MSSM at the one-loop level is
substantial in the class of MSSM scenarios with a light right stop 
\cite{window}.
Temperature loop corrections from the stop and other additional scalar
states could be large and lead to the first
order phase transition, the intensity of the latter
depends on $\xi = v(T_c)/T_c$, where $v(T_c)=\sqrt{v^2_1(T_c)+v^2_2(T_c)}$ 
is the vacuum 
expectation
value at the critical temperature $T_c$. The electroweak baryogenesis 
could be
explained if $v(T_c)/T_c >$ 1 \cite{first_criteria}, the case of strong 
first
order phase transition. 

In a number of analyses the MSSM finite-temperature
effective potential is taken in the representation
\begin{equation}
V_{eff}(v,T) = V_0(v_1,v_2,0)+V_1 (m(v),0)+V_1 (T)+V_{ring}(T),
\label{veff}
\end{equation}
where $V_0$ is the tree-level MSSM two-doublet potential at the SUSY
scale,
$V_1$ is the (non-temperature) one-loop resumed Coleman-Weinberg term, 
dominated by stop and sbottom contributions,
$V_1 (T)$ is the one-loop temperature term and $V_{ring}$ is the
correction of re-summed leading infrared contribution from multi-loop ring
(or daisy) diagrams. The MSSM relations between the $SU(2)_L \times 
U(1)_Y$
gauge couplings $g_2$ and $g_1$, and the quartic parameters 
$\lambda_{1,2,3,4}$ of the potential
$V_0(v_1,v_2,0)$ are very restrictive.
Only two additional parameters ${\tt tg}\beta=v_2/v_1$ and $m_{H^\pm}$
(charged scalar mass) 
determine the zero-temperature two-doublet Higgs sector at tree-level. The 
one-loop
radiative corrections, both logarithmic and non-logarithmic generated at 
the threshold $M_{SUSY}$, can change strongly the tree-level picture.
They depend on the parameters ($A_{t,b}$, $\mu$, $m_Q$, $m_U$,
$m_D$) of the scalar quarks-Higgs bosons interaction sector. In most 
cases for the analysis in the
representation (\ref{veff})
numerical methods are used to find the critical
temperature $T_c$, for example, by solving the equation for the 
determinant of second derivatives of the potential (\ref{veff}) at 
$v_{1,2}=$0 \cite{brignole}. Then the two background fields $v_{1,2} 
(T_c)$ are found at 
the minimum using the minimization conditions (i.e. the absence of linear 
terms of the effective potential representation in the "shifted" fields). 
The first order phase transition strength is dependent on the 
cubic term $ETv^3$ which appears from the infrared region.

Numerical high-precision Monte Carlo simulations on the lattice 
\cite{montecarlo} 
have been developed and applied to MSSM in connection with the infrared 
problem \cite{irproblem} inherent to all analyses based on the effective 
potentials. Infrared divergences appear in the integration over bosonic static 
($\omega_0=$0) Matsubara modes, 
which in the loop expansion for the three-dimensional momentum space 
correspond to the intermediate massless bosons. The 
non-perturbative investigations of the problem have been performed in the 
framework of high-temperature dimensional reduction 
\cite{KajLaiRumSha1995,dimred}, when an 
effective three-dimensional MSSM with the same Green's functions as in 
the four-dimensional MSSM for the light bosons is constructed 
\cite{laine, losada, cline} by integrating out perturbatively the 
non-static modes. The corrections from squarks and gauge bosons are 
introduced after the reduction to the three-dimensional model.

In order to cover the temperature range from very low temperatures to 
the temperatures of the order of critical, the following analysis uses
an approach developed in \cite{physrev, echaya} 
for the general (non-temperature) two-Higgs doublet potential with
complex-valued parameters $\mu^2_{12}$, $\lambda_5$, $\lambda_6$ and 
$\lambda_7$, which violates the CP-invariance explicitly. 
However, in this publication a simplified situation of the Higgs potential 
in the CP conserving limit is considered
(the imaginary parts of the effective parameters $\lambda_{5,6,7}$ and 
$\mu^2_{12}$ are taken to be zero). Full MSSM  
effective potential in the generic $\Phi_1$, $\Phi_2$ basis has the form
\begin{equation}
U_{eff}(\Phi_1,\Phi_2) =
- \, \mu_1^2 (\Phi_1^\dagger\Phi_1) - \, \mu_2^2 (\Phi_2^\dagger
\Phi_2) - \mu_{12}^2 (\Phi_1^\dagger \Phi_2) -
\stackrel{*}{\mu_{12}^2} (\Phi_2^\dagger \Phi_1) 
 + \lambda_1
(\Phi_1^\dagger \Phi_1)^2
     +\lambda_2 (\Phi_2^\dagger \Phi_2)^2
\label{eq:genU}
\end{equation}
$$ + \lambda_3 (\Phi_1^\dagger \Phi_1)(\Phi_2^\dagger \Phi_2) +
\lambda_4 (\Phi_1^\dagger \Phi_2)(\Phi_2^\dagger \Phi_1) 
+ \frac{\lambda_5}{2}
(\Phi_1^\dagger \Phi_2)(\Phi_1^\dagger \Phi_2) + 
\frac{\lambda^*_5}{2}
(\Phi_2^\dagger \Phi_1)(\Phi_2^\dagger \Phi_1) +
$$
$$
 + \lambda_6
(\Phi^\dagger_1 \Phi_1)(\Phi^\dagger_1 \Phi_2)+
\stackrel{*}{\lambda}_6(\Phi^\dagger_1 \Phi_1)(\Phi^\dagger_2
\Phi_1) + \lambda_7 (\Phi^\dagger_2 \Phi_2)(\Phi^\dagger_1 \Phi_2)
+\stackrel{*}{\lambda}_7(\Phi^\dagger_2 \Phi_2)(\Phi^\dagger_2
\Phi_1)  $$
\vskip 2mm \noindent
where the background fields (vev's) are
$\langle \Phi_1 \rangle=(0,v_1)/\sqrt{2}$ and
$\langle \Phi_2 \rangle=(0,v_2)/\sqrt{2}$.
The temperature corrections from squarks, both logarithmic and 
non-logarithmic (at the SUSY threshold) are incorporated
to $\lambda_1,...\lambda_7$.
In \cite{physrev, echaya} (see also \cite{other})
a nonlinear transformation for masses and mixing angles
$\lambda_i=\lambda_i(\alpha,\beta,m_h,m_H,m_A,m_{H^\pm} 
\lambda_6,\lambda_7)$, $ \, i=1,...5$ to the Higgs bosons mass basis can be 
found for a general case ($h$,$H$ and $A$ are the neutral and $H^+$, $H^-$ are 
the charged Higgs bosons, $\alpha$ is the $h$-$H$ mixing angle, 
${\tt tg}\beta=v_2/v_1$)
\begin{equation}
U_{eff}(\Phi_1,\Phi_2)
\Longrightarrow \frac{m^2_h}{2} (hh) + \frac{m^2_H}{2} (HH) +
\frac{m^2_A}{2} (AA) + m^2_{H^\pm} (H^+ H^-)
+ h,H,A,H^\pm \quad \mbox{interaction terms}
\label{masseig}
\end{equation}
which allows to work with symbolic expressions for the temperature-dependent Higgs boson mass 
eigenstates. 

In section 2 various one-loop temperature corrections to the
potential are calculated. Section 3 contains some examples of the electroweak phase 
transition for the finite-temperature effective potential reconstructed in the full MSSM parameter space. The potential of scalar quarks - Higgs bosons interaction and some technical details of evaluation can be found in the Appendix. 


\section{Finite temperature corrections of squarks}
\vskip 0.2cm

\noindent
In the finite temperature field theory Feynman diagrams with boson
propagators, containing Matsubara frequencies $\omega_n=2\pi n T$
($n=0,\pm1,\pm2,...$), lead to structures of the form
\begin{equation}
I[m_{1},m_{2},...,m_{b}]= T \sum_{n=-\infty}^{\infty}\int\frac{d {\bf k}}
{(2\pi)^{3}}
\prod_{i=1}^{b}\frac{(-1)^b}{({\bf k}^2+\omega_n^2+m_j^2)},
\label{commonI}
\end{equation}
Here $\bf k$ is the three-dimensional momentum in a system with the
temperature $T$. In the following calculations first we perform integration with respect to
$\bf k$ and then take the sum, using the reduction to three-dimensional
theory in the high-temperature limit for zero frequencies. At $n\not=0$
the result is~\cite{LV97,Amore2005}
\begin{equation}
I[m_{1},m_{2},...,m_{b}] = 2 T \left(2\pi T\right)^{3-2 b} \
\frac{(-1)^b\pi^{3/2}}{(2\pi)^3} \
\frac{\Gamma(b-3/2)}{\Gamma(b)} \ S(M,b-3/2),
\label{pr case}
\end{equation}
where
\begin{equation}
S(M,b-3/2)= \int \{dx\} \,\sum_{n=1}^\infty \ \frac{1}{(n^2+M^2)^{b-3/2}},
\qquad M^2 \equiv \left( \frac{m}{2 \pi T}\right)^2.
\label{sum}
\end{equation}
For $b>1$ the parameter $m^2$ is a linear function dependent on $m_i^2$
and the variables $\{dx\}$ of Feynman parametrization, which are the
integration variables in~(\ref{sum}). At the integer values of $b$
the integrand in (3) is a generalized Hurwitz zeta-function \cite{hurwitz}.
Note that for the
leading threshold corrections to effective parameters of the
two-doublet potential $b>2$, so the wave-function renormalization appears 
in
connection with the divergence at $b=2$ (which is suppressed by vertex
factors, see \cite{physrev}).

A number of integrals can be easily calculated. The integral $J_0$ is calculated
\begin{equation}
J_0[a_1,a_2] = \int\frac{d{\bf k}}{(2\pi)^3}\frac{1}{({\bf
k}^2+a_1^2)({\bf k}^2+a_2^2)} =
\frac{1}{4\pi (a_1+a_2)},
\label{commonI2}
\end{equation}
taking a residue in the spherical coordinate system.
Here $a_{1;2}^2$ are the sums of squared frequency and squared mass,
see~(\ref{commonI}). Derivatives of $J_0$ with respect to $a_1$ and $a_2$
can be used for calculation of integrals
\begin{equation}
J_1[a_1,a_2]= \int\frac{d{\bf k}}{(2\pi)^3}\frac{1}
{({\bf k}^2+a_1^2)^2({\bf k}^2+a_2^2)}= \, - \, \frac{1}{2
a_1}\frac{\partial J_0}
{\partial a_1}=
\frac{1}{8\pi a_1(a_1+a_2)^2},
\label{intI1}
\end{equation}
\begin{equation}
J_2[a_1,a_2]=
\int\frac{d{\bf k}}{(2\pi)^3}\frac{1}{({\bf
k}^2+a_1^2)^2({\bf k}^2+a_2^2)^2}=
\frac{1}{4 a_1 a_2}\frac{\partial^2 J_0}{\partial a_1 \partial a_2}=
\frac{1}{8\pi a_1 a_2(a_1+a_2)^3}.
\label{intI2}
\end{equation}
and
\footnote{The same results for $J_3$ and $J_4$ can be found in
\cite{laine} and \cite{losada}, where they appear in the context of high
temperature dimensional reduction. }
\begin{eqnarray}
J_3[a_1,a_2,a_3] = \int\frac{d {\bf k}}{(2\pi)^3}
\frac{1}{({\bf k}^2+a_1^2)({\bf k}^2+a_2^2)({\bf k}^2+a_3^2)} = \frac{1}{4\pi
(a_1+a_2)(a_1+a_3)(a_2+a_3)},
\label{L}
\end{eqnarray}
\begin{eqnarray}
J_4[a_1,a_2,a_3] = \int\frac{d {\bf k}}{(2\pi)^3}
\frac{1}{({\bf k}^2+a_1^2)^2({\bf k}^2+a_2^2)({\bf k}^2+a_3^2)}=
\frac{2a_1+a_2+a_3}{8\pi
a_1 (a_1+a_2)^2(a_1+a_3)^2(a_2+a_3)},
\label{L1}
\end{eqnarray}
Thus, the procedure of Feynman parametrization is not
used. Substitution of $a^2_1 = {4 \pi^2 n^2 T^2+m_1^2}$ and $a^2_2 =
{4 \pi^2 n^2 T^2+m_2^2}$ to (\ref{commonI2})
and summation over Matsubara frequencies after the integration
gives
\begin{equation}
I_0[m_1,m_2]=\sum_{n=-\infty, n \neq
0}^{\infty} J^n_0 [m_1,m_2] =\sum_{n=-\infty, n \neq 0}^{\infty}\frac{1}
{4\pi (\sqrt{4 \pi^2 n^2 T^2+m_1^2}+\sqrt{4
\pi^2 n^2 T^2+m_2^2})}.
\label{diversum}
\end{equation}
or, after redefinition of mass parameters $M_{1;2}={m_{1;2}}/{2\pi
T}$ the temperature corrections to effective potential are expressed 
by summed integrals
\begin{equation}
 I_{1}[M_1,M_2] =-\frac{ 1}{64 \pi^4
T^{2}}\sum_{n=-\infty, n \neq 0}^{\infty}
\frac{1}{\sqrt{M_{1}^{2}+n^{2}}
(\sqrt{M_{1}^{2}+n^{2}}+\sqrt{M_{2}^{2}+n^{2}})^{2}},
\label{Ias} \end{equation}
\begin{equation}
I_{2}[M_1,M_2] =\frac{1}{256 \pi^5
T^{4}}\sum_{n=-\infty, n \neq 0}^{\infty}
\frac{1}{\sqrt{M_{1}^{2}+n^{2}}\sqrt{M_{2}^{2}+n^{2}}
(\sqrt{M_{1}^{2}+n^{2}}+\sqrt{M_{2}^{2}+n^{2}})^{3}}.
\label{Iabs}
\end{equation}
Note that the series
(\ref{diversum}) are divergent, but the derivatives (\ref{Ias}) and
(\ref{Iabs}) are convergent for all $M_{1;2}$. In the following it will be
convenient to keep separately terms for zero and nonzero modes 
in the
sum. Both terms will be temperature-dependent since the zero-mode
integrals coincide with (\ref{commonI2})-(\ref{intI2}), where
$a_i^2=m_i^2$ and the factor $T$ should be accounted for. Numerical check
of the
zero temperature limiting case $T \rightarrow 0$ demonstrates that the
non-temperature field theory results are successfully
reproduced. In the high-temperature limit the zero mode gives dominant
contribution in agreement with a known suppression of quantum effects
at increasing temperatures.

The sum of integrals (\ref{Ias}) and (\ref{Iabs}) can be expressed by
means
of the generalized zeta-function. Such forms can be derived if we
introduce Feynman parameters in the integrand of (\ref{commonI2})
\begin{equation}
\label{Ifp}
\frac{1}{[{\bf k}^{2}+m_{a}^{2}][{\bf k}^{2}+m_{b}^{2}]}=
\int_{0}^{1}\frac{dx}{([{\bf
k}^{2}+m_{a}^{2}]x+[{\bf k}^{2}+m_{b}^{2}](1-x))^{2}},
\end{equation}
and redefine ${\bf k}\longrightarrow {\bf p}={\bf k}/{2\pi T}$,
$M^2(M_{a}, M_{b}, x) = (M_{a}^{2}-M_{b}^{2})x+M_{b}^{2}$.
Then we get
\begin{equation}
\label{Ifp2}
\frac{1}{[{\bf k}^{2}+m_{a}^{2}][{\bf k}^{2}+m_{b}^{2}]}=\frac{1}{(2 \pi
T)^{4}}\int_{0}^{1}\frac{dx}{({\bf p}^{2}+n^{2}+M^{2})^{2}}.
\end{equation}
and divergent series for (\ref{commonI2}) ($d{\bf k}=(2 \pi T)^{3}
d{\bf p}$)
\begin{equation}
\label{I3}
I_0[M_a, M_b] = \frac{1}{2 \pi T}\int_{0}^{1}dx
\sum_{n=-\infty, n \neq
0}^{\infty}\int
\frac{d{\bf p}}{(2\pi)^{3}}\frac{1}{({\bf p}^{2}+n^{2}+M^{2})^{2}},
\end{equation}
With the help of dimensional regularization or differentiating the
integral
\begin{equation}
\label{I4}
\int\frac{d{\bf p}}{(2\pi)^{3}}\frac{1}{({\bf p}^{2}+M^{2})}
= -\frac{M}{4\pi}+{\cal O}(\frac{M^2}{T^2})
\end{equation}
over the parameter $M$, the equation (\ref{I3}) can be reduced to
\begin{equation}
\label{I5} I_0[M_a, M_b] =\frac{1}{16 \pi^{2} T }\int_{0}^{1}dx \;
\zeta(2, \frac{1}{2}, M^{2}),
\end{equation}
where $\zeta(u, s, t)$ is the generalized Hurwitz zeta-function
\cite{hurwitz}:
\footnote{Note that (non-generalized) Hurwitz zeta-function is defined by
$\zeta(s, t)=\sum_{n=0}^{\infty}\frac{1}{[n+t]^{s}}$.}
\begin{equation}
\label{zeta}
\zeta(u, s, t)=\sum_{n=1}^{\infty}\frac{1}{(n^{u}+t)^{s}}.
\end{equation}
So in the case under consideration the sums of integrals (\ref{Ias}) and 
(\ref{Iabs}) can be calculated by differentiation of (\ref{I5}) with respect to mass
parameters participating in $M=M(M_{a}, M_{b}, x)$. Differentiation
increases the power $s$ in the denominator of (\ref{I5}) giving convergent
integrals
\begin{equation}
\label{Ias2}
I_{1}[M_a, M_b] =\frac{T}{2M_{a}}\frac{\partial}{\partial M_{a}}
I_0=-\frac{1}{64\pi^4 T^{2}}\int_{0}^{1} \; dx \; x \;
\zeta[2, \frac{3}{2}, M^{2}(x)],
\end{equation}
\begin{equation}
\label{Iabs2}
I_{2}[M_a, M_b] =-\frac{1}{2M_{b}}\frac{\partial}{\partial M_{b}}
(-I_{1})=\frac{3}{256 \pi^6 T^{4}}\int_{0}^{1}dx \; x \;
(1-x) \; \zeta[2, \frac{5}{2}, M^{2}(x)].
\end{equation}
The integrals (\ref{Ias2}) and (\ref{Iabs2}) are equal to the
series (\ref{Ias}) and (\ref{Iabs}), respectively.
\unitlength 1.00mm
\begin{figure}[h]
\begin{center}
\begin{picture}(40,40)
\put(-100,-205){\epsfxsize=21cm \epsfysize=27cm \leavevmode
\epsfbox{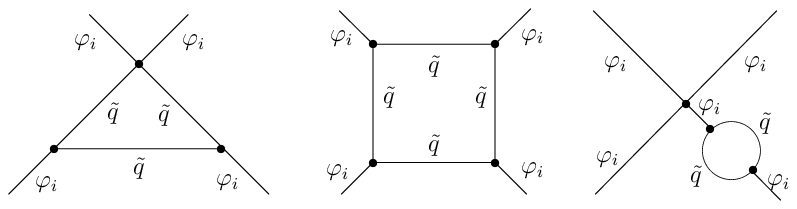}}
\end{picture}
\end{center}
\label{thresh}
\vspace{-4mm}
\caption{\small
Threshold corrections (left and central diagram) and diagram 
contributing to the wave-function renormalization (right).}
\end{figure}

{\bf Threshold corrections from the triangle and box diagrams}, 
shown in Fig.\ref{thresh}, are denoted by
$\Delta\lambda^{th}_i$, $i$=1,...7.
They contribute additively to the 
parameters $\lambda_i=\lambda^{SUSY}_i-\Delta\lambda^{th}_i$. 
In the following 
normalization conventions from \cite{physrev} are used. 
Calculation of the finite-temperature diagrams for the general case 
of complex-valued $\mu$ and $A_{t,b}$ gives the result
(see details in the Appendix)
\begin{equation}
\Delta\lambda^{thr}_1=3 h_t^4 |\mu|^4 I_2[m_Q,m_U]
+3 h_b^4 |A_b|^4 I_2[m_Q,m_D]+
\label{eq:lambda1}
\end{equation}
$$+h_t^2|\mu|^2
( - \, \frac{g_1^2-3g_2^2}{2}I_1[m_Q,m_U]+2 g_1^2 I_1[m_U,m_Q]) $$ 
$$+h_b^2|A_b|^2 (\frac{12h_b^2-g_1^2-3g_2^2}{2}I_1[m_Q,m_D]+(6h_b^2-
g_1^2) I_1[m_D,m_Q])$$

\begin{equation}\Delta\lambda^{thr}_2=3 h_t^4 |A_t|^4 I_2[m_Q,m_U]
+3 h_b^4 |\mu|^4 I_2[m_Q,m_D]+\end{equation}$$+h_b^2|\mu|^2
(\frac{g_1^2+3g_2^2}{2}I_1[m_Q,m_D]+ g_1^2 I_1[m_D,m_Q]) +$$ $$+
h_t^2|A_t|^2 (\frac{12h_t^2+g_1^2-3g_2^2}{2}I_1[m_Q,m_U]+(6h_t^2- 2
g_1^2) I_1[m_U,m_Q])$$

\begin{equation}\Delta\lambda^{thr}_3=h_t^2((|\mu|^2 
\frac{3g_2^2+g_1^2}{12}+
|A_t|^2\frac{12h_t^2-g_1^2-3g_2^2}{12}) I_1[m_Q,m_U]+\end{equation}
$$+(|\mu|^2 \frac{3h_t^2-g_1^2}{3}+
|A_t|^2\frac{g_1^2}{3}) I_1[m_U,m_Q])+$$
$$+(h_b^2(|\mu|^2 \frac{3g_2^2-g_1^2}{12}+
|A_b|^2\frac{12h_t^2+g_1^2-3g_2^2}{4}) I_1[m_Q,m_D]+$$
$$+(|\mu|^2 \frac{6h_b^2-g_1^2}{6}+
|A_b|^2\frac{g_1^2}{6}) I_1[m_D,m_Q])+$$
$$
+h_t^2|\mu|^2|A_t|^2I_2[m_Q,m_U]+h_b^2|\mu|^2|A_b|^2I_2[m_Q,m_D]+
$$
$$+h_t^2h_b^2(2(A_t
A_b-|\mu|^2)I_3[m_Q,m_U,m_D]+(|\mu|^4+|A_t|^2|A_b|^2-2A_t
A_b|\mu|^2)I_4[m_Q,m_U,m_D]$$
\begin{equation}
\Delta\lambda^{thr}_4=6 h_t^4 |\mu|^2 |A_t|^2 I_2[m_Q,m_U]
+6 h_b^4 |\mu|^2 |A_b|^2 I_2[m_Q,m_D]+\end{equation}$$+h_t^2((|\mu|^2
\frac{12h_t^2+g_1^2-3g_2^2}{4}-|A_t|^2\frac{g_1^2-3g_2^2}{4})
I_1[m_Q,m_U]+$$
$$+(|A_t|^2 g_1^2 -|\mu|^2(g_1^2-3h_t^2))I_1[m_U,m_Q]) +$$
$$+ h_b^2((|\mu|^2
\frac{-12h_t^2+g_1^2+3g_2^2}{4}-|A_b|^2\frac{g_1^2+3g_2^2}{4})
I_1[m_Q,m_D]+$$
$$+\frac{1}{2}(|A_b|^2 g_1^2
-|\mu|^2(g_1^2-6h_b^2))I_1[m_D,m_Q])-\Delta\lambda^{th}_3$$
\begin{equation}
\label{eq:lambda5}
\Delta\lambda^{thr}_5=3 h_t^4 \mu^2 A_t^2 I_2[m_Q,m_U]
+3 h_b^4 \mu^2 A_b^2 I_2[m_Q,m_D]
\end{equation}
\begin{equation}
\label{eq:lambda6}
\Delta\lambda^{thr}_6=-3 h_t^4 \mu A_t |\mu|^2 I_2[m_Q,m_U]
-3 h_b^4 \mu A_b |A_b|^2 I_2[m_Q,m_D]+
\end{equation}$$+h_t^2 \mu A_t
(\frac{g_1^2-3g_2^2}{4}I_1[m_Q,m_U]- g_1^2 I_1[m_U,m_Q]) +$$ $$+
h_b^2 \mu A_b
(\frac{-12h_b^2+g_1^2+3g_2^2}{4}I_1[m_Q,m_D]-\frac{6h_b^2- g_1^2}{2}
I_1[m_D,m_Q])$$
\begin{equation}
\Delta\lambda^{thr}_7=-3 h_t^4 \mu A_t |A_t|^2 I_2[m_Q,m_U]
-3 h_b^4 \mu A_b |\mu|^2  I_2[m_Q,m_D]
\label{eq:lambda7}
\end{equation}$$+h_b^2 \mu A_b
(-\frac{g_1^2+3g_2^2}{4}I_1[m_Q,m_D]- \frac{g_1^2}{2} I_1[m_D,m_Q])
+$$ $$+ h_t^2 \mu A_t
(\frac{12h_t^2+g_1^2-3g_2^2}{4}I_1[m_Q,m_U]-(3h_t^2- g_1^2)
I_1[m_U,m_Q])$$
where $g_1$, $g_2$ are U(1) and SU(2) gauge couplings, $\mu$ is the Higgs
superfield mass parameter, $A_t$, $A_b$ are the trilinear 
squarks-Higgs bosons parameters,
$h_{\,t}, h_{\,b}$ are the Yukawa couplings and $m_Q, m_U, m_D$ denote the
scalar quark mass parameters, in terms of which the physical masses are expressed.
\unitlength 1.00mm
\begin{figure}[h]
\begin{center}
\begin{picture}(40,40)
\put(-100,-205){\epsfxsize=21cm \epsfysize=27cm \leavevmode
\epsfbox{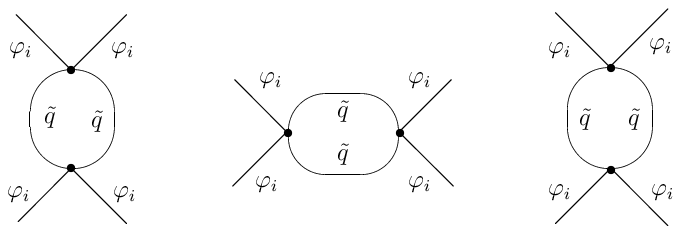}}
\end{picture}
\end{center}
\vspace{-10mm}
\caption{\small "Fish" diagrams \label{fishdiag}}
\end{figure}

{\bf Corrections of "fish" diagrams}, see Fig.\ref{fishdiag}, give the 
following contributions to the effective parameters
\begin{equation} 
- \Delta\lambda_1^{f} =
\left[h_b^2-\frac{g_1^2}{6}\right]^2(I(m_Q) + I(m_D)) +
\frac{g_1^4}{9}I(m_U),
\label{fishfirsteq}
\end{equation}
\begin{equation} 
- \Delta\lambda_2^{f} = \left[h_t^2
+\frac{g_1^2}{6}\right]^2I(m_Q) +
[h_t^2-\frac{g_1^2}{3}]^2I(m_U)+\frac{g_1^4}{36}I(m_D),
\end{equation}

$$ - (\Delta\lambda_3+\Delta\lambda_4)^{f} = \frac{1}{72} \left(-g_1^4+6
   (h_b^2-h_t^2)g_1^2-9( g_2^4 - 2 ( h_b^2+h_t^2)
g_2^2)\right) I(m_Q)+$$ 
\begin{equation}
+\frac{g_1^2}{3}(h_t^2-\frac{g_1^2}{3})I(m_U)
+\frac{g_1^2}{6}(h_b^2-\frac{g_1^2}{6})I(m_D),
\end{equation}

$$ - \Delta\lambda_3^{f} = \frac{1}{72} \left(-g_1^4+6
   (h_b^2-h_t^2)g_1^2+9
   \left(g_2^4-2\left(h_b^2+h_t^2\right)
   g_2^2+8 h_b^2
   h_t^2\right)\right)I(m_Q)+$$
\begin{equation}
+\frac{g_1^2}{3}(h_t^2-\frac{g_1^2}{3})I(m_U)
+\frac{g_1^2}{6}(h_b^2-\frac{g_1^2}{6})I(m_D)+h_t^2h_b^2I(m_U,m_D).
\end{equation}

\begin{equation}
- \Delta\lambda_4^{f}
=(h_b^2-\frac{g_2^2}{2})(\frac{g_2^2}{2}-h_t^2)I(m_Q)-h_t^2h_b^2I(m_U,m_D).
\label{fishlasteq}
\end{equation}

The three-dimensional integrals in (\ref{fishfirsteq})-(\ref{fishlasteq})
are 
\begin{equation} 
J(m_I) = \frac{1}{8\pi m_I}, \qquad J(m_U,m_D) = \frac{1}{4\pi (m_U+m_D)}.
\end{equation} 
see (\ref{commonI2}), leading to series analogously to (\ref{diversum}) 
and (\ref{I3}).

{\bf The logarithmic corrections} for non-degenerate squark masses can be 
defined 
following \cite{haberhempf} and \cite{choidreeslee}. Schematically, in 
the 
results of \cite{echaya}) we replace 
$\ln \left(\frac{M_{SUSY}}{m_t^2}\right)$ by
$\ln \left(\frac{m_Q m_U}{m_t^2}\right)$ :
\begin{equation}
\Delta\lambda_1^{log}=
-\frac{1}{384 \pi ^2}\left(11 g_1^4-36 h_b^2 g_1^2+9 \left(g_2^4-4 h_b^2 
g_2^2
+16h_b^4\right)\right) \ln \left(\frac{m_Q m_U}{m_t^2}\right),
\end{equation}
\begin{equation}
\Delta\lambda_2^{log}=-\frac{1}{1536 \pi ^2}\left(44 g_1^4-144 h_t^2 g_1^2
+36 g_2^4+576 h_t^4-144g_2^2 h_t^2\right)
\ln \left(\frac{m_Q m_U}{m_t^2}\right),
\end{equation}
\begin{equation}
\Delta\lambda_3^{log}=-\frac{1}{384 \pi ^2}\left(-11 g_1^4
+18 \left(h_b^2+h_t^2\right) g_1^2+\right.
\end{equation}
$$\left.+9 \left(g_2^4-2
   \left(h_b^2+h_t^2\right) g_2^2+16 h_b^2 h_t^2\right)\right)
\ln \left(\frac{m_Q m_U}{m_t^2}\right),$$
\begin{equation}
\Delta\lambda_4^{log}=\frac{3}{64 \pi ^2} \left(g_2^4-2 \left(h_b^2
+h_t^2\right) g_2^2+8 h_b^2 h_t^2\right)
\ln \left(\frac{m_Q m_U}{m_t^2}\right).
\end{equation}
Large logarithms not connected with the renormalization group appear
also in the wave-function renormalization yield, see below.

It is known that in order to renormalize the $\lambda \varphi^4$ 
theory, one needs to renormalize the self-coupling and the mass of the 
scalar 
field. If the $\lambda \varphi^4$ theory is supplemented by fermions 
with interactions defined by the Yukawa term, an additional wave-function 
renormalization is necessary. Similar situation takes place in the 
two-doublet model. Expanding the self-energy diagram (see the insertion 
to the leg in Fig.\ref{thresh}, right) calculated with non-degenerate masses 
at finite temperature, we get at $p^2=$0 {\bf the wave-function 
renormalization (w.f.r.) correction}, which is defined by a factor in 
front of $p^2$. At 
zero temperature two ways of w.f.r. calculation can be used 
\cite{collins}. The following calculation is based on the integration of convergent 
w.f.r. contribution over the momentum squared, previously which has been 
used in differentiation. The standard subtraction scheme at zero momentum 
(BPSZ-scheme) in the divergent expression for the self-energy 
contribution, when the divergent pole part is subtracted, turns out to be 
not convenient at finite temperatures, because in summation over Matsubara 
frequencies not divergent integrals, but divergent series must be 
subtracted. Following \cite{physrev} we can write
\begin{equation} \Delta\,\lambda_{\,1}^{\rm wfr} = 
\frac{1}{2} (g_1^2+g_2^2)
A'_{11} , \qquad \Delta\,\lambda_{\,2}^{\rm wfr} = \frac{1}{2}
(g_1^2+g_2^2) A'_{22} , \label{eq:frc} \end{equation}
$$ \Delta\,\lambda_{\,3}^{\rm wfr} = -\,\frac{1}{4}
(g_1^2-g_2^2) (A'_{11}+A'_{22}) , \quad \Delta\,\lambda_{\,4}^{\rm
wfr} = -\,\frac{1}{2} g_2^2 (A'_{11}+A'_{22}) , \quad
\Delta\,\lambda_{\,5}^{\rm wfr}=0,$$
$$ \Delta\,\lambda_{\,6}^{\rm wfr}  = \frac{1}{8} (g_1^2+g_2^2)
(A'_{12}-{A'_{21}}^*) = 0, \quad \Delta\,\lambda_{\,7}^{\rm wfr} =
\frac{1}{8} (g_1^2+g_2^2) (A'_{21}-{A'_{12}}^*) = 0 . $$
where $A$ matrices 2$\times$2 are
\footnote{The equations in \cite{physrev} are given for the general case 
of complex-valued $\mu$, $A_{t,b}$.}
\begin{equation} A'_{\,ij} =\{\frac{2\cdot3h_U^2}{24\,\pi}
F(m^{2}_{{Q}},m^{2}_{{U}},T)\left[\begin{array}{cc} |\mu|^2 &
- \mu^* A_U^*\\ - \mu A_U & |A_U|^2 \end{array}\right] \,
+({U}\longrightarrow {D}, A\longleftrightarrow \mu)\}(1-\frac{1}{2}l),
 \label{eq:fieldbezlogcor}
\end{equation}
include the series
(compare with Eq.(113) in~\cite{KajLaiRumSha1995}, taking into
account differentiation to get the finite w.f.r. yield)
\begin{equation}
F(m^{2}_{1},m^{2}_{2},T)=T\sum^{+\infty}_{n=-\infty}
\frac{1}{(\sqrt{m^{2}_{1}+(2\pi
nT)^{2}}+\sqrt{m^{2}_{2}+(2\pi nT)^{2}})^{3}}=
\end{equation}
$$=\frac{T}{(m_{1}+m_{2})^{3}}+2T\sum^{+\infty}_{n=1}
\frac{1}{(\sqrt{m^{2}_{1}+(2\pi nT)^{2}}+\sqrt{m^{2}_{2}
+(2\pi nT)^{2}})^{3}}.$$
The sum of all w.f.r. corrections to $\lambda_{5,6,7}$ vanishes.

It is useful to check that the finite temperature corrections are reduced 
to the structures of zero-temperature MSSM, which play a role of 
{\bf boundary condition at $T$=0}. Indeed 
in the limiting case of {$T$=0} and degenerate squark mass
parameters all equal to $M_{SUSY}$ the 
threshold corrections given by Eq.(\ref{eq:lambda1})-(\ref{eq:lambda7})  
are reduced to previous zero-temperature results \cite{physrev,yaflast}. 
For example, let us take $\Delta \lambda_1$ for $m_a=m_b=M_{SUSY}$
\begin{equation}\Delta\lambda_1=3 h_t^4 |\mu|^4 I_2[M_{SUSY}]
+3 h_b^4 |A|^4 I_2[M_{SUSY}]+\end{equation}$$+h_t^2|\mu|^2
(\frac{g_1^2-3g_2^2}{2}I_1[M_{SUSY}]+2 g_1^2 I_1[M_{SUSY}]) +$$ $$+
h_b^2|A|^2 (\frac{12h_b^2-g_1^2-3g_2^2}{2}I_1[M_{SUSY}]+(6h_b^2-
g_1^2) I_1[M_{SUSY}]),$$
where the integrals are
\begin{equation}
I_1[M_{SUSY}]\equiv-\int\frac{d^4k}{(2\pi)^4}
\frac{1}{(k^2+M_{SUSY}^2)^3}=
-\frac{1}{16\pi^2} \frac{1}{2M_{SUSY}^2},
\label{I1T0susy}
\end{equation}
\begin{equation}
I_2[M_{SUSY}]\equiv\int\frac{d^4k}{(2\pi)^4}\frac{1}{(k^2+M_{SUSY}^2)^4}=
\frac{1}{16\pi^2} \frac{1}{6M_{SUSY}^4}
\label{I2T0susy}
\end{equation}
Transformation to Minkowski space leads to the change of sign in
(\ref{I1T0susy}). The equality of the temperature series for $I_{1,2}$ to 
the symbolic expressions for the integrals can be numerically verified.

In the limiting case of {$T$=0} and different squark mass
parameters the reduction of (\ref{Ias}) and (\ref{Iabs}) to
the four-dimensional $I_{1}$ and $I_{2}$ can be achieved using
\begin{equation}
\label{Ia0pr}
\frac{1}{[p^{2}+m_{a}^{2}]^{2}[p^{2}+m_{b}^{2}]}=
-\frac{1}{2m_{a}}\frac{\partial}{\partial
m_{a}}\frac{1}{[p^{2}+m_{a}^{2}][p^{2}+m_{b}^{2}]},
\end{equation}
Differentiating (\ref{Ia0pr}) with respect to $m_{b}$
\begin{equation}
\label{Iab0pr}
\frac{1}{[p^{2}+m_{a}^{2}]^{2}[p^{2}+m_{b}^{2}]^{2}}=
-\frac{1}{2m_{b}}\frac{\partial}{\partial
m_{b}}\frac{1}{[p^{2}+m_{a}^{2}]^{2}[p^{2}+m_{b}^{2}]}.
\end{equation}
then using Feynman parametrization (\ref{Ifp}), differentiation in the same way as in
(\ref{Ia0pr}) and (\ref{Iab0pr}), and dimensional regularization to
integrate over the four-momentum $p$ with the following integration
over the Feynman parameter, we arrive at
\begin{equation}
\label{Ia4D}
I_{1}= - \int\frac{d^{4}p}{(2\pi)^{4}}
\frac{1}{[p^{2}+m_{a}^{2}]^{2}[p^{2}+m_{b}^{2}]}=
-\frac{m_{a}^{2}-m_{b}^{2}(1+2ln\frac{m_{a}}{m_{b}})}
{16\pi^{2}(m_{a}^{2}-m_{b}^{2})^{2}},
\end{equation}
\begin{equation}
\label{Iab4D}
I_{2}=\int\frac{d^{4}p}{(2\pi)^{4}}\frac{1}
{[p^{2}+m_{a}^{2}]^{2}[p^{2}+m_{b}^{2}]^{2}}=
\frac{m_{a}^{2}-m_{b}^{2}-(m_{a}^{2}+m_{b}^{2})
ln\frac{m_{a}}{m_{b}}}{8\pi^{2}(m_{a}^{2}-m_{b}^{2})^{3}}.
\end{equation}

\unitlength 1.00cm
\begin{figure}[h!]
\begin{center}
\begin{picture}(8,6)
\put(-1.5,-0.5){\epsfxsize=9cm \epsfysize=7cm \leavevmode
\epsfbox{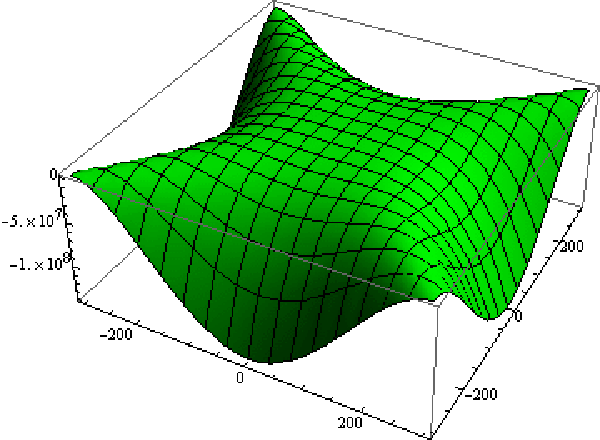}}
\end{picture}
\end{center}
\caption{\small
The zero-temperature surface of extrema for the two-doublet Higgs potential $U_0 (v_1,v_2)$,
see (\ref{eq:genU}), at the scale $M_{SUSY}$.}
\label{susy_potential}
\end{figure}

In the limit $m_a=m_b$ these formulas coincide with the expressions for
degenerate squark masses (\ref{I1T0susy}) and (\ref{I2T0susy}).

In calculations of the temperature dependent parameters $\lambda_i (T)$ of
the effective MSSM potential at moderate temperatures truncated series with fifty
terms (50 Matsubara frequencies) were used. Relative contributions of the
remaining terms are less than 10$^{-2}$ percent at $T$=50 GeV,
decreasing with an increasing $T$. At small temperatures of the order of a few GeV
an acceptable accuracy is achieved with 1000 terms. The effective
parameters $\lambda_i (T)$ are less than one, justifying the perturbative
approach, as a rule, at the squark mass parameters around several hundred 
GeV.
However, strong parametric dependence is observed here,
for example, at the squark mass parameters 200, 500 and 800 GeV the
criteria $\lambda_i (T) < $1 is valid up to $T \sim$ 860 GeV, while taking
degenerate squark masses at 600 GeV we found that at $T >$ 600 GeV the
perturbative regime cannot be used.


\section{Thermal evolution and the critical temperature}
\vskip 0.2cm

In view of the effective two-doublet potential structure defined by (\ref{eq:genU})
one could assume that the two-dimensional picture of a broken symmetry of $U_{eff}(v_1,v_2)$ with a local minima
at $T=$0, $v_{1,2} \neq$0 appears in the
sum of the potential terms with $\mu^2_1$, $\mu^2_2$ and $\mu^2_{12}$ of dimension 2 in the fields, which form a 'saddle' (a hyperboloid in the $(v_1,v_2)$ space), and of the dimension 4 terms
$\lambda_{1,...7}$ which are increasing quartically, being unbounded from above.
However, the situation is more involved because $\mu^2_{1}$ ,$\mu^2_{2}$, $\mu^2_{12}$ and $\lambda_i$
respect a number of constraints.
In this section we are going to describe roughly some possible scenarios 
of temperature 
evolution in the effective two-doublet MSSM Higgs sector with threshold, 
logarithmic and wave-function renormalization one-loop corrections. Two sets of the 
squark mass parameters in the following numerical calculations are used

\noindent
(A) $m_Q=$500 GeV, $m_U=$200 GeV, $m_D=$800 GeV,\\
(B) $m_Q=$500 GeV, $m_U=$800 GeV, $m_D=$200 GeV.

Masses of the third generation squarks are
\begin{eqnarray*}
m^2_{\tilde{t}_{1,2}}=\frac{1}{2}((m^2_{\tilde{t}_L}+m^2_{\tilde{t}_R})
                   \mp \sqrt{(m^2_{\tilde{t}_L}-m^2_{\tilde{t}_R})^2+4 \, A^2_1 \, m^2_{top}}),\\
m^2_{\tilde{b}_{1,2}}=\frac{1}{2}((m^2_{\tilde{b}_L}+m^2_{\tilde{b}_R})
                   \mp \sqrt{(m^2_{\tilde{b}_L}-m^2_{\tilde{b}_R})^2+4 \, A^2_2 \, m^2_{b}}),\\
\end{eqnarray*}
where 
\begin{eqnarray*}
m^2_{\tilde{t}_L}=m^2_Q+m^2_{top}+{\tt cos} \, 2\beta \, m^2_Z \, (\frac{1}{2}
                                                     -\frac{2}{3} {\tt sin} \theta^2_{w}),\\
m^2_{\tilde{t}_R}=m^2_U+m^2_{top}+{\tt cos} \, 2\beta \, m^2_Z \, 
                                                     (\frac{2}{3} {\tt sin} \theta^2_{w}),\\
m^2_{\tilde{b}_L}=m^2_Q+m^2_{b}+{\tt cos} \, 2\beta \, m^2_Z \, (-\frac{1}{2}
                                                     +\frac{1}{3} {\tt sin} \theta^2_{w}),\\
m^2_{\tilde{b}_R}=m^2_D+m^2_{b}+{\tt cos} \, 2\beta \, m^2_Z \, (
                                                     -\frac{1}{3} {\tt sin} \theta^2_{w})
\end{eqnarray*}

\newpage
\unitlength 1.00cm
\begin{figure}[h!]
\begin{center}
\begin{picture}(8,18.0)
\put(-5.5,15.2){\epsfxsize=8cm \epsfysize=4.3cm \leavevmode
\epsfbox{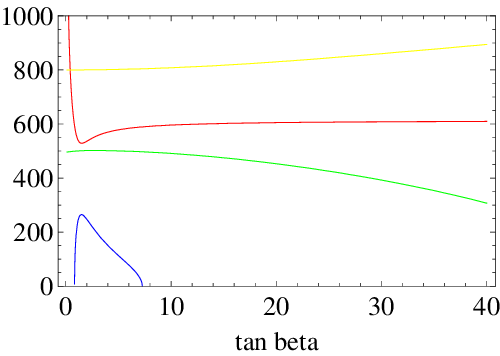}}
\put(-3.5,16.4){\mbox{$m_{{\tilde t}_1}$}}
\put(-3.5,18.2){\mbox{$m_{{\tilde t}_2}$}}
\put(-3.5,17.3){\mbox{$m_{{\tilde b}_1}$}}
\put(-3.5,19.0){\mbox{$m_{{\tilde b}_2}$}}
\put(4.0, 15.2){\epsfxsize=8cm \epsfysize=4.3cm \leavevmode
\epsfbox{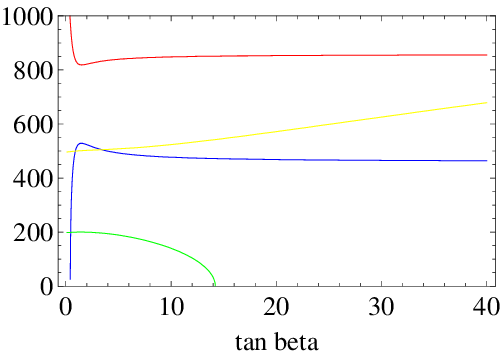}}
\put(7.2,16.4){\mbox{$m_{{\tilde b}_1}$}}
\put(7.2,18.2){\mbox{$m_{{\tilde t}_1}$}}
\put(7.2,17.3){\mbox{$m_{{\tilde b}_2}$}}
\put(7.2,19.0){\mbox{$m_{{\tilde t}_2}$}}
\put(-5.5,10.6){\epsfxsize=8cm \epsfysize=4.3cm \leavevmode
\epsfbox{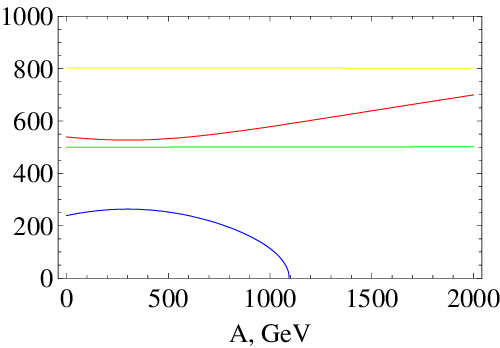}}
\put(-3.5,11.7){\mbox{$m_{{\tilde t}_1}$}}
\put(-3.5,13.5){\mbox{$m_{{\tilde t}_2}$}}
\put(-3.5,12.6){\mbox{$m_{{\tilde b}_1}$}}
\put(-3.5,14.3){\mbox{$m_{{\tilde b}_2}$}}
\put(4.0, 10.6){\epsfxsize=8cm \epsfysize=4.3cm \leavevmode
\epsfbox{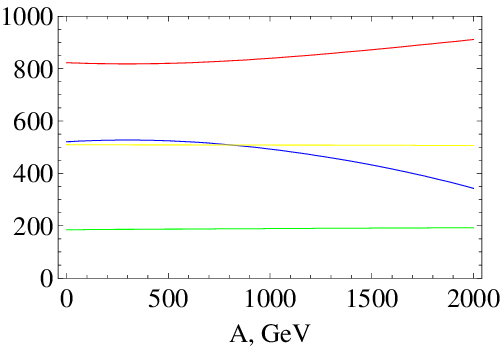}}
\put(7.2,11.7){\mbox{$m_{{\tilde b}_1}$}}
\put(8.6,13.5){\mbox{$m_{{\tilde t}_1}$}}
\put(8.2,12.6){\mbox{$m_{{\tilde b}_2}$}}
\put(7.2,14.3){\mbox{$m_{{\tilde t}_2}$}}
\put(-5.5,6.1){\epsfxsize=8cm \epsfysize=4.3cm \leavevmode
\epsfbox{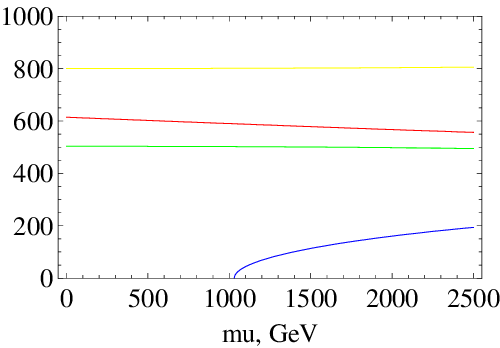}}
\put(-3.5,7.2){\mbox{$m_{{\tilde t}_1}$}}
\put(-3.5,9.0){\mbox{$m_{{\tilde t}_2}$}}
\put(-3.5,8.1){\mbox{$m_{{\tilde b}_1}$}}
\put(-3.5,9.8){\mbox{$m_{{\tilde b}_2}$}}
\put(4.0, 6.1){\epsfxsize=8cm \epsfysize=4.3cm \leavevmode
\epsfbox{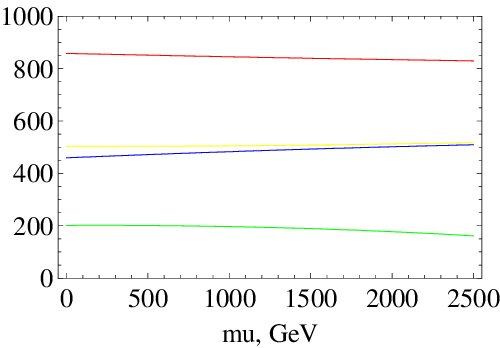}}
\put(7.2,7.2){\mbox{$m_{{\tilde b}_1}$}}
\put(5.6,8.8){\mbox{$m_{{\tilde t}_1}$}}
\put(5.6,7.9){\mbox{$m_{{\tilde b}_2}$}}
\put(7.2,9.8){\mbox{$m_{{\tilde t}_2}$}}
\put(-5.5,-0.8){\epsfxsize=8cm \epsfysize=6.8cm \leavevmode
\epsfbox{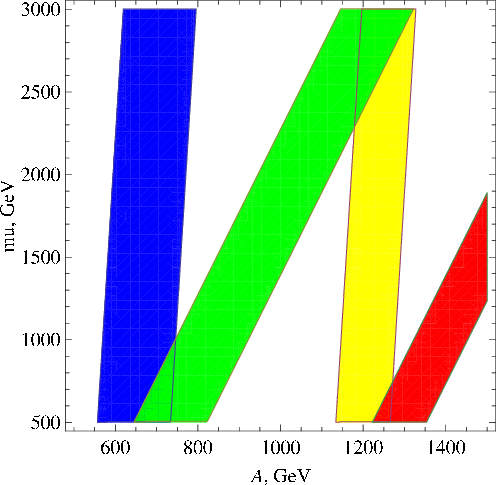}}
\put(-3.4,1.8){\mbox{(1)}}
\put(-2.0,1.8){\mbox{(2)}}
\put( 0.2,1.8){\mbox{(5)}}
\put( 1.7,1.8){\mbox{(6)}}
\put(4.0, -0.8){\epsfxsize=8cm \epsfysize=6.8cm \leavevmode
\epsfbox{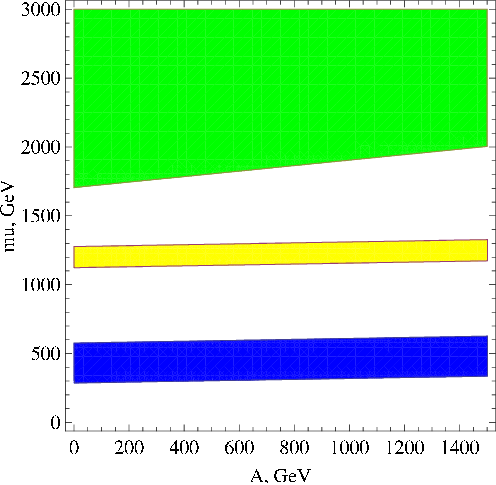}}
\put(7.2,1.0){\mbox{(4)}}
\put(7.2,2.2){\mbox{(7)}}
\put(7.2,3.9){\mbox{(3)}}
\end{picture}
\end{center}
\caption{\small
Third generation squark masses as a function of ${\tt tg}\beta$ (first row of plots), $A$ (second row) and $\mu$ (third row of plots). For the left column of plots the squark sector parameter values are $m_Q=$500 GeV, $m_U=$200 GeV, $m_D=$800 GeV, set (A). For the right column of plots the squark sector parameter values are $m_Q=$500 GeV, $m_U=$800 GeV, $m_D=$200 GeV, set (B). The fourth row of plots demonstrates the regions of "light stop" 120 GeV $<$ $m_{\tilde{t}_1}$ $<$ 180 GeV (left panel) and "light sbottom" 120 GeV $<$ $m_{\tilde{b}_1}$ $<$ 180 GeV (right panel) in the ($A$, $\mu$) plane. In the regions (1), ${\tt tg}\beta=$40 and (2), ${\tt tg}\beta=$5, squark parameters are given by set (A), in the region (3), where ${\tt tg}\beta=$5, and (4), where ${\tt tg}\beta=$30, squark parameters given by set (B). For regions (5) and (6) single parameter $m_U$ of the set (A) is shifted to 400 GeV, for region (7) single parameter $m_D$ of the set (B) is shifted to 400 GeV, other kept fixed. }
\label{mt1_mb1}
\end{figure}
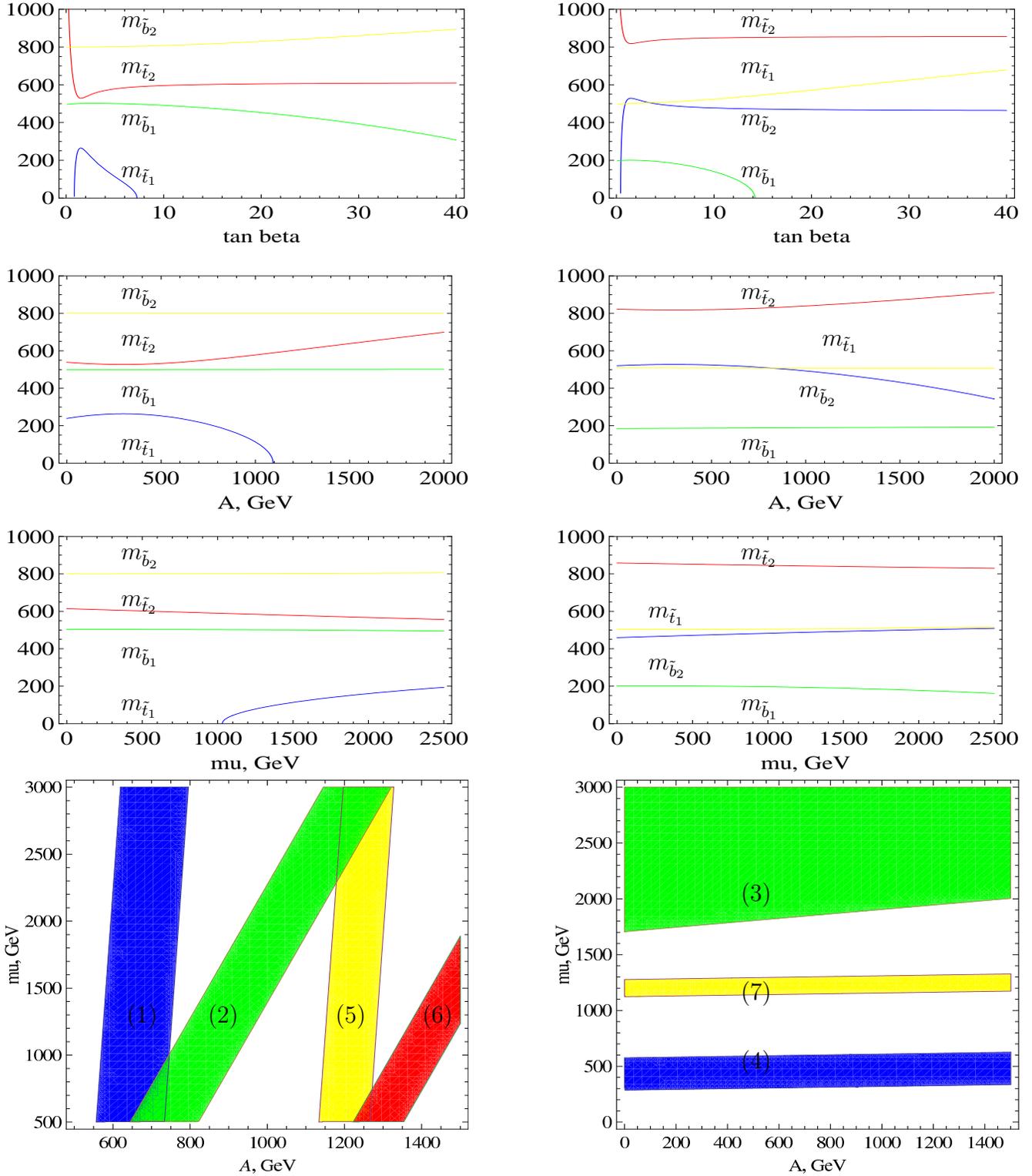
and
\begin{eqnarray*}
A^2_1=A^2_{t,b}+\mu^2 \, {\tt ctg}^2 \beta - 2\, A_{t,b}\, \mu\, {\tt ctg}\beta, \hspace{ 10mm}
A^2_2=A^2_{t,b}+\mu^2 \, {\tt tg}^2 \beta - 2\, A_{t,b}\, \mu\, {\tt tg}\beta.
\end{eqnarray*}
With these parameters the third generation squark 
eigenstates $m_{\tilde{t}_{1,2}}$ and $m_{\tilde{b}_{1,2}}$ which masses are positively defined exist, as a rule, in an extensive regions of the  (${\tt tg}\beta$, $A$, $\mu$ ) parameter space, see Fig.\ref{mt1_mb1}. Set (A) favors the light stop, while the light sbottom is a feature of the parameter set (B). Fixed parameters for the plots in Fig.\ref{mt1_mb1} are ${\tt tg}\beta=$5, $A_{t,b}=$1 TeV, $\mu=$1.5 TeV, $m_{H^\pm}=$180 GeV. Squark masses vary in the range from 200 GeV to 800 GeV at the values of $A_{t,b}$ and $\mu$ up to the order of 1 TeV . Large difference of the stop 
masses is necessary to respect constraints following from the LEP2 experimental limit $m_h >$115 GeV.
The values of ${\tt tg\beta}$ above 5 and large soft supersymmetry 
breaking parameters $A_{t,b}$ and $\mu$ of the order of $m_Q$ also lead to 
an acceptable Higgs boson mass $m_h$ (but weaken the strength of the 
electroweak phase transition if taken too large). At the same 
time substantial threshold corrections 
appear in the MSSM scenarios with large $A_{t,b}$ and $\mu$, like the BGX
scenario \cite{bgx} and the CPX scenario \cite{cpx}, or the regions of 
MSSM parameter space close to BGX and CPX.
\unitlength 1.00cm
\begin{figure}[h!]
\begin{center}
\begin{picture}(8,6)
\put(-5.5,-0.5){\epsfxsize=9cm \epsfysize=7cm \leavevmode
\epsfbox{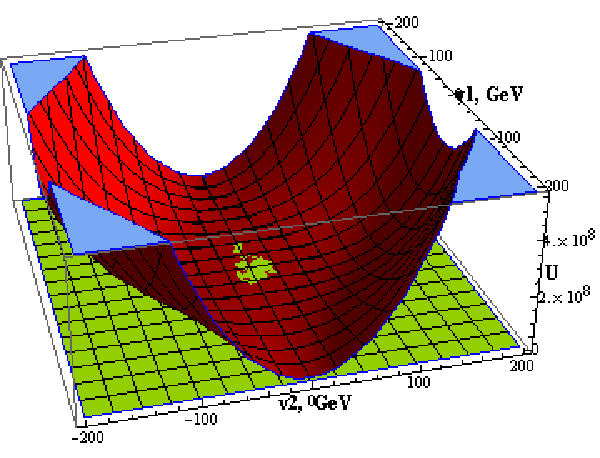}}
\put(4.0,-0.5){\epsfxsize=9cm \epsfysize=7cm \leavevmode
\epsfbox{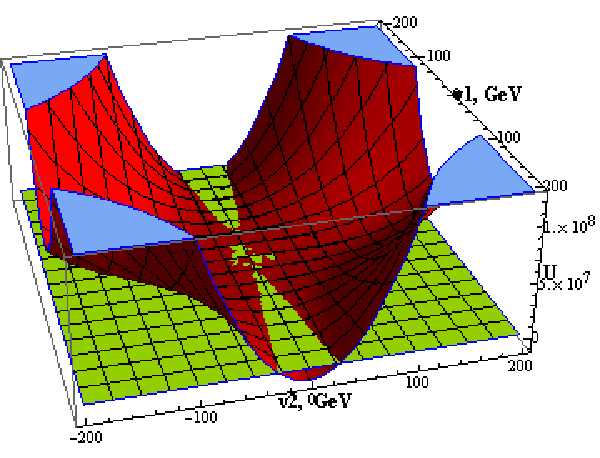}}
\end{picture}
\end{center}
\caption{\small
Development of the saddle configuration for the surface of stationary points of the potential $U_{eff}
(v_1,v_2)$, see (\ref{eq:genU}), at the critical temperature $T_c=$120
GeV. The squark sector parameter values are $m_Q=$500 GeV, $m_U=$200 GeV,
$m_D=$800 GeV, $A_t=A_b=$1200 GeV, $\mu=$500 GeV,
the charged Higgs boson mass $m_{H^\pm}=$150 GeV. Horizontal plane corresponds to $U_{eff}=$0.  }
\label{Ueff67zero}
\end{figure}

In the following the equilibrium states of the effective potential
(\ref{eq:genU}) as a function of the two variables of state $v_1$ and $v_2$ and six temperature-dependent
control parameters $\lambda_1 (T),... \lambda_7 (T)$ are going to be analyzed. Local properties of $U_{eff}(v_1,v_2,\lambda_1,...\lambda_7)$
are defined by a number of well-known theorems in the framework of the catastrophe theory (Morse and Thom theorems for the reduction of a potential function to the canonical form by a nonlinear transformation \cite{catastrophe}). They describe
properties of the stationary state $\nabla U_{eff} (v_1, v_2)=$0 defined by the stability matrix
(also called the Hessian) $U_{ij}=\partial^2 U_{eff}/ \partial v_i \partial v_j$. Simplest two-dimensional example of the 
Hessian is given by the MSSM Higgs potential at the SUSY scale, where 
$\lambda_1=\lambda_2=(g_1^2+g_2^2)/8$, 
$\lambda_3=(g_2^2-g_1^2)/4$, $\lambda_4=-g_2^2/2$ and 
$\lambda_5 =\lambda_6 =\lambda_7 =0$ are independent of the temperature. The equilibrium matrix at the stationary points has the form
\begin{equation}
\left\Vert \begin{array}{cc}
\frac{1}{4}(g^2_1+g^2_2)v^2_1+m^2_A\frac{v^2_2}{v^2_1+v^2_2} & -\frac{1}{4}(g^2_1+g^2_2)v_1 v_2-m^2_A\frac{v_1 v_2}{v^2_1+v^2_2} \\
-\frac{1}{4}(g^2_1+g^2_2)v_1 v_2-m^2_A\frac{v_1 v_2}{v^2_1+v^2_2} & \frac{1}{4}(g^2_1+g^2_2)v^2_2+m^2_A\frac{v^2_1}{v^2_1+v^2_2}
\end{array} \right\Vert 
\end{equation} 
($m_A$ is the CP-odd scalar mass)
and the nonisolated (or degenerate) critical points defined by the condition ${\tt det} \, U_{ij}=$0 lead to the equation
$(g^2_1+g^2_2) m^2_A (v^2_1 - v^2_2)^2/(v^2_1+v^2_2)=$0, so the MSSM surface of minima $U_0 (v_1,v_2)=-(g_1^2+g_2^2)(v_1^2-v_2^2)^2/32$
is unbounded from below and the bifurcation set looks as the two 'flat directions' $v_1= \pm v_2$, see 
Fig.\ref{susy_potential}. Threshold corrections at zero temperature can be 
found in \cite{yaflast}. As a rule they transform the decreasing function
in Fig.\ref{susy_potential} to a saddle configuration, slowly increasing
along one of the 'flat directions' and more rapidly decreasing along the
other. 

In the general case the potential (\ref{eq:genU}) as a function of the vacuum expectation values
\begin{equation}
U(v_1, v_2) = -\frac{\mu^2_1}{2} v^2_1 -\frac{\mu^2_2}{2} v^2_2 -\mu^2_{12} v_1 v_2
+ \frac{\lambda_1}{4}v^4_1 + \frac{\lambda_2}{4}v^4_2 + \frac{\lambda_{345}}{4}v^2_1 v^2_2
+ \frac{\lambda_6}{2}v^3_1 v_2 + \frac{\lambda_7}{2}v_1 v^3_2
\label{Uv1v2}
\end{equation}
includes temperature-dependent
parameters $\lambda_i (T)$, $i$=1,...7, and $v_{1,2} (T)$,
see Eq.(\ref{eq:lambda1})-(\ref{eq:lambda7}),
which define the thermal evolution from some high temperature $T$ of 
the order of several hundred GeV down to zero. We denote $\lambda_{345} = \lambda_{3} +\lambda_{4}+{\tt Re} \lambda_{5}$. Conditions of the extremum $\nabla U(v_1, v_2)=0$ distinguishing an isolated (or nondegenerate) critical points     
\begin{eqnarray}
\label{diagmu1}
\mu^2_{1}& = &
 \lambda_1 v^2_1+( \lambda_3+
\lambda_4+{\tt Re} \lambda_5)\frac{v^2_2}{2}
- {\tt Re}\mu_{12}^2 {\tt tg}\beta
+\frac{v^2 s^2_{\beta}}{2}
(3 {\tt Re} \lambda_6 {\tt ctg} \beta + {\tt Re} \lambda_7 {\tt tg}
\beta) , \\
\mu^2_{2}& = &  {\hskip -2mm}
 \lambda_2 v^2_2+( \lambda_3+
\lambda_4+{\tt Re} \lambda_5)\frac{v^2_1}{2} - {\tt Re}\mu_{12}^2
{\tt ctg}\beta +\frac{v^2 c^2_{\beta}}{2} ( {\tt Re} \lambda_6
{\tt ctg}\beta + 3  {\tt Re} \lambda_7 {\tt tg} \beta ) , 
\label{diagmu2}
\end{eqnarray}
where 
\begin{eqnarray*}
{\tt Re} \mu^2_{12}&=&\sin {\beta}
\cos {\beta}[m^2_A+\frac{v^2}{2}(2 {\tt Re}\lambda_5+{\tt Re}
\lambda_6 {\tt ctg} \beta+{\tt Re} \lambda_7 {\tt tg}
\beta)] ,
\end{eqnarray*}
are also mentioned as the minimization conditions which set to 
zero the linear terms in the physical fields $h$, $H$ and $A$ and ensure a 
local extremum at any point of the surface $U_{eff} (v_1,v_2)$ in the 
background fields space (see e.g. \cite{physrev})
\footnote{Although only the CP-conserving limit is considered, we keep the 
notation of real parts for the variables where a phase factor could appear 
in the general case.} 
Important input parameters of the two-doublet potential are 
${\tt tg}\beta=v_2/v_1$ and the charged 
Higgs boson mass
\begin{equation}
m^2_{H^\pm}=m^2_W+m^2_A-\frac{v^2}{2}( {\tt Re} \Delta \lambda_5-
 \Delta\lambda_4)
\end{equation}
where the effective temperature-dependent mass of the longitudinal
$W$-boson is $m^2_{W_L}(v,T)=m^2_W(v)+\Pi_{W_L}(T)$, $\Pi_{W_L}(T)=5g^2_2 
T^2/2$ (with the one-loop Standard Model and third-generation squarks 
contributions included in the polarization operator; $m^2_W=v^2 g^2_2/2$). 
If in the process of thermal evolution, when the system moves along
some trajectory in the $v_1(T), v_2(T)$ plane, we require the
minimization of $U$ with respect to the scalar fields oscillation in
the extremum $v_1(T), v_2(T)$ and continuously admit the
interpretation of the system in terms of scalar states $h, H$ and
$A$, then $\mu_1^2$, $\mu_2^2$ and $\mu_{12}^4$ can be
expressed by means of the effective parameters $\lambda_{1,...7}$ 
\cite{physrev}.
\footnote{ The normalization of $\lambda_{1,2}$ in \cite{physrev} is
different from \cite{echaya} by a factor of 2}
Only $\mu^2_{1}$, $\mu^2_2$ and $\mu^2_{12}$ are dependent on the direction in
the $(v_1, v_2)$ plane, while $\lambda_{1,...7}$
are not. 

First it is useful consider the simplified 
case $\lambda_6=\lambda_7=0$. The two-doublet Higgs potential without $\lambda_6$ 
and $\lambda_7$ terms has been considered in the context of discrete Peccei-Quinn
symmetry \cite{PQsymm}. Nonisolated (or degenerate) critical points in the $v_1, v_2$ plane,
defined by the condition ${\tt det} \, \partial^2 U/\partial v_i \partial v_j =$0, or 
\begin{equation}
det \left\Vert \begin{array}{cc}
2 \lambda_1 v^2_1 + \mu^2_{12}\frac{v_2}{v_1} & -\mu^2_{12} + \lambda_{345} v_1 v_2 \\
-\mu^2_{12} + \lambda_{345} v_1 v_2 & 2 \lambda_2 v^2_2 + \mu^2_{12}\frac{v_1}{v_2}
\end{array} \right\Vert  =0
\end{equation} 
where the minimization conditions (\ref{diagmu1}) and (\ref{diagmu2}) 
(or, equivalently, the conditions for isolated points of $U(v_1, v_2)$)
have been substituted. The system of two nonlinear equations for $v_1, v_2$
\begin{eqnarray}
\label{nonlineareqs}
\lambda_1 v^3_1 + \frac{\lambda_{345}}{2} v_1 v^2_2 - \mu^2_1 v_1 - \mu^2_{12} v_2 =0 \\ \nonumber
\lambda_2 v^3_2 + \frac{\lambda_{345}}{2} v^2_1 v_2 - \mu^2_2 v_2 - \mu^2_{12} v_1 =0 
\end{eqnarray}
can be factorized by the rotation in the $v_1, v_2$ plane
\begin{equation}
v_1 = \bar{v}_1 {\tt cos}{\bar \beta} - \bar{v}_2 {\tt sin}{\bar \beta}, \hskip 6mm
v_2 = \bar{v}_1 {\tt sin}{\bar \beta} + \bar{v}_2 {\tt cos}{\bar \beta}
\label{betarotation}                
\end{equation}
where
\begin{equation}
{\tt sin}^2{\bar \beta}=
   \frac{1}{2} \pm \frac{|\mu^2_1-\mu^2_2|}{2 \sqrt{(\mu^2_1-\mu^2_2)^2+4\mu^4_{12}}}, \hskip 6mm
{\tt cos}^2{\bar \beta}=
   \frac{1}{2} \mp \frac{|\mu^2_1-\mu^2_2|}{2 \sqrt{(\mu^2_1-\mu^2_2)^2+4\mu^4_{12}}}
\label{anglebeta}
\end{equation}
Then the factorized equations (\ref{nonlineareqs}) are
\begin{eqnarray}
\bar{v}_1 (\lambda_1 \bar{v}_1^2 + \frac{\lambda_{345}}{2} \bar{v}_2^2 - \bar{\mu}_1^2)=0 \\ \nonumber
\bar{v}_2 (\lambda_2 \bar{v}_2^2 + \frac{\lambda_{345}}{2} \bar{v}_1^2 - \bar{\mu}_2^2)=0
\end{eqnarray}
where
\begin{equation}
{\bar \mu}^2_{1,2}= \frac{1}{2}(\mu^2_1+\mu^2_2 \pm \sqrt{(\mu^2_1-\mu^2_2)^2 + 4\mu^4_{12}})
\end{equation}
and the four types of bifurcation sets defined by the 
stability matrices $U_{ij}(v_1, v_2)$ can be easily found

(1) $\lambda_1 \bar{v}_1^2 + \frac{\lambda_{345}}{2} \bar{v}_2^2 - \bar{\mu}_1^2=$0 and
 $\lambda_2 \bar{v}_2^2 + \frac{\lambda_{345}}{2} \bar{v}_1^2 - \bar{\mu}_2^2=$0,
$U_{ij}(v_1,v_2)= \left\Vert \begin{array}{cc} 
                    2 \lambda_1 \bar{v}_1^2 & \lambda_{345} \bar{v}_1 \bar{v}_2 \\
                     \lambda_{345} \bar{v}_1 \bar{v}_2 &  2 \lambda_2 \bar{v}_2^2
                \end{array} \right\Vert $

\vskip 3mm
(2) $\lambda_1 \bar{v}_1^2 - \bar{\mu}_1^2=$0 and $\bar{v}_2=$0,
$U_{ij}(v_1,v_2)= \left\Vert \begin{array}{cc} 
                    2 \lambda_1 \bar{v}_1^2 & 0 \\
                     0 &  -\bar{\mu}_2^2 + \frac{\lambda_{345}}{2} \bar{v}_1^2
                \end{array} \right\Vert $
\vskip 3mm
(3) $\bar{v}_1=$0 and $\lambda_2 \bar{v}_2^2 - \bar{\mu}_2^2=$0,
$U_{ij}(v_1,v_2)= \left\Vert \begin{array}{cc} 
                    -\mu^2_1 + \frac{\lambda_{345}}{2} \bar{v}_2^2 & 0 \\
                     0 &  2 \lambda_2 \bar{v}_2^2
                \end{array} \right\Vert $
\vskip 3mm
(4) $\bar{v}_1=$0 and $\bar{v}_2=$0, 
          $U_{ij}(v_1,v_2)= -\left\Vert \begin{array}{cc} \bar{\mu_1}^2 & 0 \\
                                                          0 & \bar{\mu_2}^2 
                                         \end{array} \right\Vert $
\unitlength 1.00cm
\begin{figure}[h!]
\begin{center}
\begin{picture}(8,6)
\put(-4.,-0.5){\epsfxsize=7cm \epsfysize=7cm \leavevmode
\epsfbox{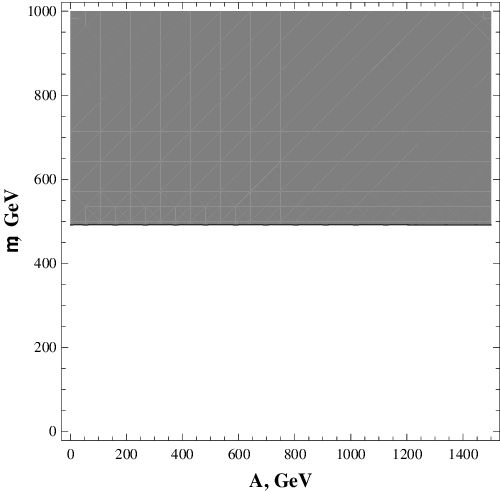}}
\put(4.5,-0.5){\epsfxsize=7cm \epsfysize=7cm \leavevmode
\epsfbox{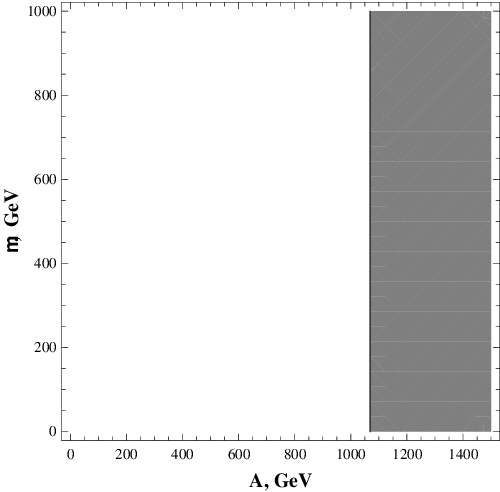}}
\end{picture}
\end{center}
\caption{\small
Contours of negatively defined $\lambda_1$ (left, dark grey
area) and $\lambda_2$ (right, dark grey area) in the ($A_t=A_b$, $\mu$)
plane at the temperature 150 GeV, $m_{H^\pm}=$150 GeV. Set (A), the case of light stop, is used for the squark sector parameter values
($m_Q=$500 GeV, $m_U=$200 GeV,
$m_D=$800 GeV).}
\label{lambdas12}
\end{figure}

\unitlength 1.00cm
\begin{figure}[h!]
\begin{center}
\begin{picture}(8,6)
\put(-0.5,-0.5){\epsfxsize=7cm \epsfysize=7cm \leavevmode
\epsfbox{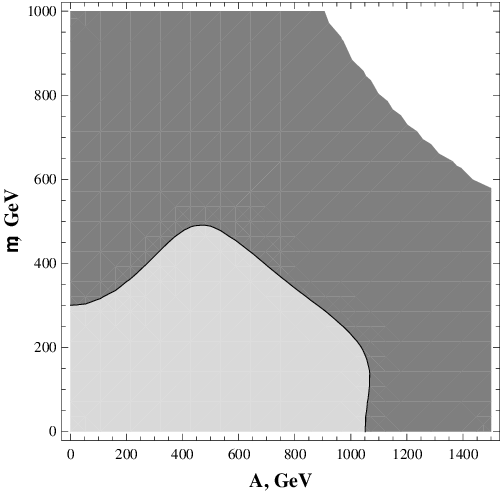}}
\end{picture}
\end{center}
\caption{\small
Contour of negatively defined determinant $\lambda_1
\lambda_2-\lambda^2_{345}/4$ (dark grey area) in the ($A_t=A_b$, $\mu$)
plane at the temperature 150 GeV. Set (A), the case of light stop, is used for the squark sector parameter values
($m_Q=$500 GeV, $m_U=$200 GeV,
$m_D=$800 GeV).}
\label{conditionlambdas}
\end{figure}

Bifurcation set in the case (1) which is defined by ${\tt det} \, \partial^2 U/\partial v_i \partial v_j =$0
can be understood in the elementary language. The surface of stationary points 
$U_{eff}(v_1,v_2)=-(\lambda_1 v_1^4 + \lambda_2 v_2^4 + \lambda_{345} 
v_1^2 v_2^2)/4$
is positively defined and unbounded from above if the Sylvester's criteria for the 
quadratic form $U_{eff} (v_1^2, v_2^2)$ is respected
\begin{equation}
\lambda_1 < 0, \quad \lambda_2 <0, \quad 
\lambda_1 \lambda_2 - \frac{\lambda^2_{345}}{4}<0
\label{uneq}
\end{equation}
At the critical temperature defined by the equation $\lambda_1 \lambda_2 - 
\lambda^2_{345}/{4}=$0 the positively defined potential surface of stationary points starts to develop the saddle 
configuration which is unbounded from below, see Fig.\ref{Ueff67zero}. The "flat direction" at the critical temperature
which is developed at the angle ${\tt tg}\, 2 {\theta} = \lambda_{345}/(\lambda^2_1-\lambda^2_2)$, or
\begin{equation}
\label{angle1}
{\tt tg}^2 {\theta} = \frac{\lambda^2_{345}}{(|\lambda_1-\lambda_2|-\sqrt{(\lambda_1-\lambda_2)^2+\lambda^2_{345}})^2}
\label{flat1}
\end{equation}
is defined by the control parameters $\lambda_1 (T)$, $\lambda_2 (T)$ and $\lambda_{345} (T)$ not depending on the $v_1$ and $v_2$.
The regions of 
positively and negatively defined $\lambda_1$ and $\lambda_2$
and the contour for Sylvester's criteria (\ref{uneq}) 
are shown in Figs. \ref{lambdas12} and \ref{conditionlambdas}
at the temperature $T=$150 GeV in the ($A=A_{t}=A_{b}$, $\mu$) plane. 
The squark mass parameters $m_Q$, $m_U$ and 
$m_D$ are fixed as mentioned in the beginning of the section, set (A),
the ($A$,$\mu$) parameters are chosen in the vicinity of 
the contours which separate positively and negatively defined 
$\lambda$-parameters in (\ref{uneq}). The critical temperature in this 
case 
is slightly above 120 GeV, insignificantly dependent on the values of 
($A_{t,b}$, $\mu$) if they are changing
along the contours in Fig.\ref{lambdas12}-\ref{conditionlambdas}, separating the light grey 
and the dark grey areas. 
\unitlength 1.00cm
\begin{figure}[h!]
\begin{center}
\begin{picture}(8,6)
\put(-4.,-0.5){\epsfxsize=7cm \epsfysize=7cm \leavevmode
\epsfbox{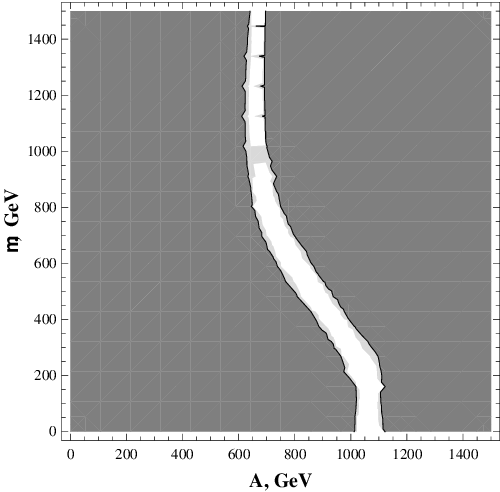}}
\put(4.5,-0.5){\epsfxsize=7cm \epsfysize=7cm \leavevmode
\epsfbox{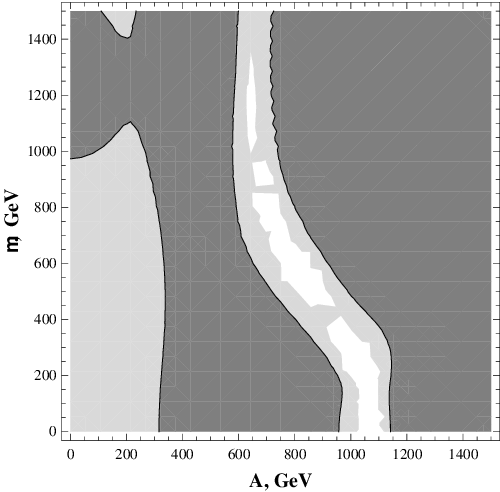}}
\end{picture}
\end{center}
\caption{\small
Contours for the criteria $\frac{v_c}{T_c} =$ 1 in the ($A_t$=$A_b$, $\mu$) plane. In the light grey regions $\frac{v_c}{T_c} >$1. 
In order to include qualitatively the effect of $E_{MSSM}$, for the left plot $E=2E_{SM}$ and for the right plot $E=4E_{SM}$.
$\lambda_6=\lambda_7=$0, charged Higgs boson mass $m_{H^\pm}=$150 GeV. Set (A), the case of light stop, is used for the squark sector parameter values
($m_Q=$500 GeV, $m_U=$200 GeV, $m_D=$800 GeV).}
\label{vctcplot}
\end{figure}
The strength of the electroweak phase transition along the direction (\ref{flat1}) can be roughly estimated using the equation 
\begin{equation}
\frac{v(T_c)}{T_c}= \frac{2\sqrt{2} \, E}{\lambda(\theta)}
\label{vctc}
\end{equation} 
where $E$ is a temperature-independent factor in front of the cubic term $-ETv^3$ in the effective potential rewritten in the polar coordinates ($v=\sqrt{v^2_1+v^2_2}$, $\theta={\tt arctan}(v_2/v_1)$ ), and $\lambda(\theta)$ is a factor in front of the quartic term 
$v^4/4$. The cubic term is given by corrections coming from the resummation of the multiloop diagrams in the infrared region. 
In the case of a heavy stop which decouples \cite{brignole}, the effective potential is similar to Standard Model potential and
\begin{equation}
E_{SM}=\frac{2\sqrt{2}}{48\pi} \, [2 g^3_2 + (g^2_1+g^2_2)^{3/2}]= \frac{\sqrt{2}}{3} \frac{(2m^3_W+m^3_Z)}{\pi v^3}.
\end{equation}
In the case of a light stop one can use an approximation $E=E_{SM}+E_{MSSM}$, where an additional term \cite{window} 
\begin{equation}
E_{MSSM}=\frac{2\sqrt{2}}{3\pi v^3} m^3_t \, (1-\frac{\tilde{A}^2_t}{m^2_Q})^{\frac{3}{2}},
\end{equation}
stop mixing parameter here $\tilde{A}_t=A_t-\mu/{\tt tg} \beta$.
The quartic term along the direction (\ref{flat1}) can be written in the form
\begin{equation}
\lambda(\theta) = -\frac{\lambda_1 + \lambda_{345} {\tt tg}^2 \theta + \lambda_{2} {\tt tg}^4 \theta + 2\lambda_6 {\tt tg} \theta
 + 2\lambda_7 {\tt tg}^3 \theta}{(1 + {\tt tg}^2 \theta)^2}.
\end{equation}
The condition $v_c/T_c>$1 \cite{first_criteria}, necessary to avoid sphaleron transitions which erase the baryon asymmetry initially generated at the electroweak phase transition, can be respected in a rather extensive regions of the ($A$,$\mu$) plane. The contours of $v_c/T_c>$1 in the
($A$,$\mu$) plane (see Fig.\ref{vctcplot}) separate the regions not only around the origin ($A$,$\mu$)=(0,0), but also the areas with ($A$,$\mu$) of the order of 1 TeV, where the quartic term $\lambda(\theta)$ changes sign crossing zero along the flat direction (\ref{flat1}).

If the set (B) is chosen for the squark mass parameters $m_Q$, $m_U$ and 
$m_D$, corresponding to the case of light sbottom and relatively heavy stop, then the factor $\lambda_2$ is always positive in the broad range of temperatures from a few to a several thousands of GeV,
the surface of stationary points is always a saddle, so the potential does not have a stable minimum at the origin $v_1=v_2=$0. 

For the general case of nonzero $\lambda_6$ and 
$\lambda_7$ defined by Eqs.(\ref{eq:lambda6}) and (\ref{eq:lambda7}) the 
effective 
potential 
$U_{eff}(v_1,v_2)=-(\lambda_1 v_1^4 + \lambda_2 v_2^4 + \lambda_{345}
v_1^2 v_2^2+2\lambda_6 v^3_1 v_2+2\lambda_7 v_1 v^3_2)/4$
always demonstrates a saddle configuration for the surface of stationary points, which slopes become steeper 
with an increase of the temperature. Typical shape of $U_{eff}(v_1,v_2)$ 
is shown in Fig.\ref{Ueff67nonzero}.
\unitlength 1.00cm
\begin{figure}[h!]
\begin{center}
\begin{picture}(8,6)
\put(-0.5,-0.5){\epsfxsize=9cm \epsfysize=7cm \leavevmode
\epsfbox{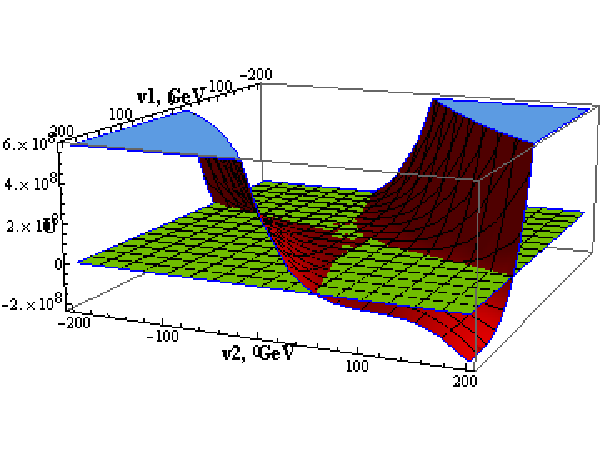}}
\end{picture}
\end{center}
\caption{\small
The surface of extrema for the potential $U_{eff} (v_1,v_2)$, see (\ref{eq:genU}), with 
nonzero $\lambda_6$ and $\lambda_7$ at the temperature $T=$120 GeV.
The squark sector parameter values $m_Q=$500 GeV, $m_U=$200 GeV,
$m_D=$800 GeV, $A_t=A_b=$1500 GeV, $\mu=$1000 GeV, charged Higgs boson 
mass $m_{H^\pm}=$150 GeV. Horizontal plane corresponds to $U_{eff}=$0.}
\label{Ueff67nonzero}
\end{figure}

Bifurcation set in the cases (2) and (3) is different from the bifurcation set in the case (1). The condition
${\bar v}_1=$0 is equivalent to 
\begin{equation}
|v_1 v_2 (\mu^2_1-\mu^2_2)|=|\mu^2_{12} (v^2_1-v^2_2)|
\end{equation}
so it follows from (\ref{anglebeta}) that ${\tt sin} {\bar \beta} = v^2_1/v^2$ and ${\tt cos} {\bar \beta} = v^2_2/v^2$ (where
$v^2=v^2_1+v^2_2$), so ${\bar v}^2_2= v^2$. Then ${\bar \mu}^2_1=(\mu^2_1+\mu^2_2+\mu^2_{12} v^2/v_1 v_2)/2$. 
The factor ${\bar \lambda}_{345}$ in the potential with rotated vacuum expectation values ${\bar U}({\bar v}_1, {\bar v}_2)$ can be found substituting (\ref{betarotation}) to (\ref{Uv1v2}) 
\begin{equation}
{\bar \lambda}_{345} = \frac{1}{4}(6\lambda_1 c^2_\beta s^2_\beta + 6\lambda_2 c^2_\beta s^2_\beta
                             + \lambda_{345} (s^4_\beta - 4 s^2_\beta c^2_\beta + c^4_\beta)) = 
(\lambda_1+\lambda_2) \frac{3 v^2_1 v^2_2}{2 v^4} + \lambda_{345} \frac{v^4_1 + v^4_2 -4 v^2_1 v^2_2}{4 v^4}
\end{equation}
Using these equations one can rewrite the conditions for the case (2) ${\bar v}_1=$0 and ${\bar \mu}^2_1={\bar \lambda}_{345} {\bar v}^2_2/2$
in the form
\begin{equation}
(4\lambda_1 + \lambda_{345}) v^4_1 + (4\lambda_2 + \lambda_{345}) v^4_2 + (6\lambda_{345} - 2\lambda_1 - 2\lambda_2) v^2_1 v^2_2 = 0
\label{uneq_2}
\end{equation}
The regions of 
positively and negatively defined $\lambda_1$ and $\lambda_2$
and the contour for Sylvester's criteria for the form (\ref{uneq_2}) 
are shown in Figs. \ref{lambdas12_2} and \ref{conditionlambdas_2}
at the temperature $T=$150 GeV in the ($A=A_{t}=A_{b}$, $\mu$) plane. 
The squark mass parameters $m_Q$, $m_U$ and 
$m_D$ are fixed as mentioned in the beginning of the section, set (A),
the ($A$,$\mu$) parameters are chosen in the vicinity of 
the contours which separate positively and negatively defined 
$\lambda$-parameters in (\ref{uneq_2}). As for the analysis of bifurcation set (1), set (B) again does not demonstrate a stable minimum at the origin for a broad range of temperatures. 
\unitlength 1.00cm
\begin{figure}[h!]
\begin{center}
\begin{picture}(8,6)
\put(-4.,-0.5){\epsfxsize=7cm \epsfysize=7cm \leavevmode
\epsfbox{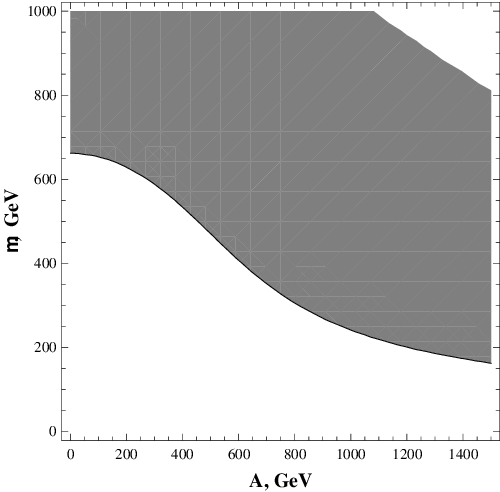}}
\put(4.5,-0.5){\epsfxsize=7cm \epsfysize=7cm \leavevmode
\epsfbox{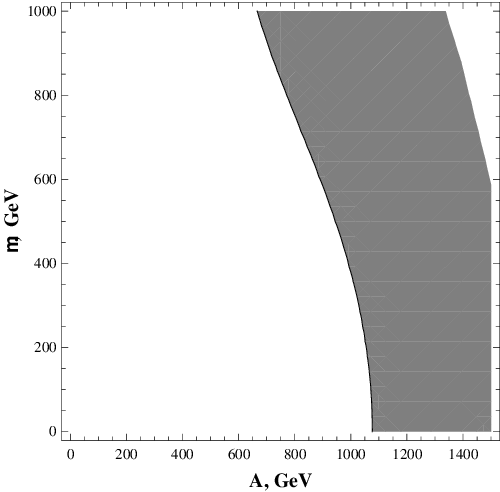}}
\end{picture}
\end{center}
\caption{\small
Contours of negatively defined $4\lambda_1+\lambda_{345}$ (left, dark grey
area) and $4\lambda_2+\lambda_{345}$ (right, dark grey area) in the ($A_t=A_b$, $\mu$)
plane at the temperature 150 GeV, $m_{H^\pm}=$150 GeV. Set (A), the case of light stop, is used for the squark sector parameter values 
($m_Q=$500 GeV, $m_U=$200 GeV,
$m_D=$800 GeV).}
\label{lambdas12_2}
\end{figure}

\unitlength 1.00cm
\begin{figure}[h!]
\begin{center}
\begin{picture}(8,6)
\put(-0.5,-0.5){\epsfxsize=7cm \epsfysize=7cm \leavevmode
\epsfbox{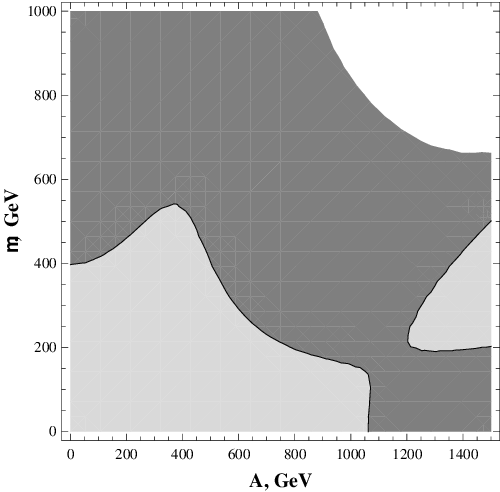}}
\end{picture}
\end{center}
\caption{\small
Contour of negatively defined determinant $(4\lambda_1+\lambda_{345})(4\lambda_2+\lambda_{345})-
(3\lambda_{345}-\lambda_1-\lambda_2)^2$ (dark grey area) in the ($A_t=A_b$, $\mu$)
plane at the temperature 150 GeV. Set (A), the case of light stop, is used for the squark sector parameter values
($m_Q=$500 GeV, $m_U=$200 GeV,
$m_D=$800 GeV).}
\label{conditionlambdas_2}
\end{figure}
The phase transition for the case (2) is developed in the direction $\theta$ of the ($v_1, v_2)$ plane
\begin{equation}
\label{angle34}
{\tt tg} 2\theta = \frac{3(3\lambda_{345}-\lambda_1-\lambda_2)}{(4\lambda_1+\lambda_{345})^2-(4\lambda_2+\lambda_{345})^2}
\end{equation}

Bifurcation set in the case (4) ${\bar v}_1=$0 and ${\bar v}_2=$0 defined by the equation ${\bar \mu}_1^2 {\bar \mu}_2^2=$0
can also be understood on the elementary level as a result of the diagonalization of the effective potential
$U_{eff}= -\frac{\mu^2_1}{2} v^2_1 -\frac{\mu^2_2}{2} v^2_2 - \mu^2_{12} v^2_1 v^2_2$ by the rotation (\ref{betarotation}),
giving the form  $U_{eff}= -{\bar \mu}^2_1 {\bar v}^2_1 -{\bar \mu}^2_2 {\bar v}^2_2$.
This case is interesting not only in the case of an effective field theory under consideration but also in a more general physical framework. So far it has been assumed that we are in the framework of an effective field theory 
at the $m_{top}$ energy scale, when the contributions from squarks decouple or a
contribution of the potential terms with squarks (see the Appendix) is practically constant.
However, if it is not the case 
and the Higgs bosons-squarks quartic term is positive definite with the global minimum 
at the origin $v_1=v_2=$0, the phase transition may occur due to the development of a 
saddle configuration by the $\mu^2_1$,$\mu^2_2$ and $\mu^2_{12}$ terms.
Such situation may take place when the vacuum expectation values of 
charged and colored superpartners participate in the full MSSM scalar 
potential, possibly giving charge and color breaking minima 
\cite{kusenko}. For illustrative purposes it is convenient 
to use the polar coordinates $v_1(T)=v(T) \cos \bar \beta(T), \quad
v_2(T)=v(T) \sin \bar \beta(T)$ for the vacuum expectation values. 
The mass term of the two-doublet potential has the form
\begin{equation}
U_{mass} (v,\bar \beta) = -\frac{v^2}{2} (\mu_1^2 \cos^2 \bar \beta
  + \mu_2^2 \sin^2 \bar \beta)  -\frac{v^2}{2} \mu_{12}^2 \sin 2\bar \beta
\label{umass}
\end{equation}
By definition at the critical temperature the gradient of $U_{mass}
(v,\bar \beta)$ is zero along some direction in the ($v_1$,$v_2$) plane,
then $\partial U_{mass}/\partial v = 0$ and $1/v \; \partial
U_{mass}/\partial \bar \beta = 0$; it follows from these two equations
\begin{equation}
{\tt tg} 2\bar \beta = \frac{2 \mu_{12}^2}{\mu_1^2-\mu_2^2}, \hspace{5mm}
(\mu_1^2 \mu_2^2 - \mu_{12}^4)[(\mu_1^2-\mu_2^2)^2+4\mu_{12}^4]=0
\label{angle}
\end{equation}
The first of these equations is equivalent to (\ref{anglebeta}).
The phase transition is characterized by the critical angle $\bar
\beta(T)$ which defines the flat direction\footnote{Flat directions may exist also in the quartic term separately 
taken, see e.g. \cite{affleck}.}
for the mass term at the temperature
$T_c$ in the background fields plane $(v_1, v_2)$, and at the real-valued
$\mu_1$, $\mu_2$ and $\mu_{12}$ the critical temperature is defined by the
equation
\begin{equation}
\mu_1^2 \mu_2^2 = \mu_{12}^4
\label{eqT}
\end{equation}
which is equivalent to ${\bar \mu}^2_1=$0 or ${\bar \mu}^2_2=$0.
For a fixed set of the squark sector parameters\footnote{{\it Mathematica} package \cite{wolfram} with encoded 
representations of $\lambda_i(T)$ by means of series with $n$=50
was used to scan the MSSM parameter space. At low temperatures the 
convergence of Matsubara series becomes worse, so the number of terms up 
to $n$=1000 is needed to reach an acceptable accuracy. } 
the thermodynamical evolution of the 
effective potential is described by a $\nabla U_{eff}=$0 trajectory in the three-dimensional
($v$, $T$, ${\tt tg}\beta$) space, which is defined by the intersection
of the two surfaces, corresponding to the equation (\ref{angle}) for the
critical angle ("$\beta$-surface"), and the equation (\ref{eqT}) for the
$\mu_1$, $\mu_2$ and $\mu_{12}$ ("$\mu$-surface"). 
The cross sections of $\mu$-surface (calculated 
without any approximations numerically) by the plane at fixed $T$=0, 
giving the ($v$, ${\tt tg}\beta$) contour, and the cross section at ${\tt 
tg}\beta=$1 giving the ($v$, $T$) contour is shown in Fig.\ref{projections}.
\unitlength 1.00cm
\begin{figure}[h!]
\begin{center}
\begin{picture}(8,6)
\put(-4.,-0.5){\epsfxsize=7cm \epsfysize=7cm \leavevmode
\epsfbox{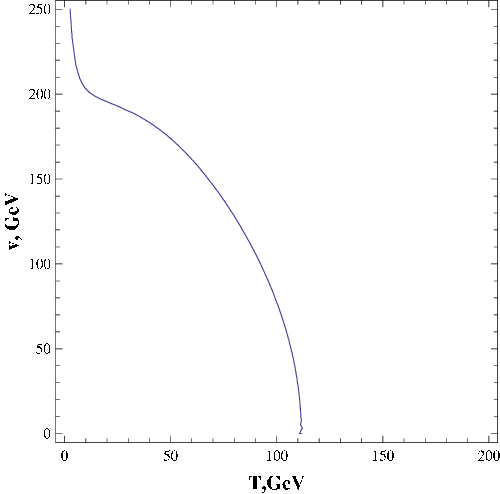}}
\put(4.5,-0.5){\epsfxsize=7cm \epsfysize=7cm \leavevmode
\epsfbox{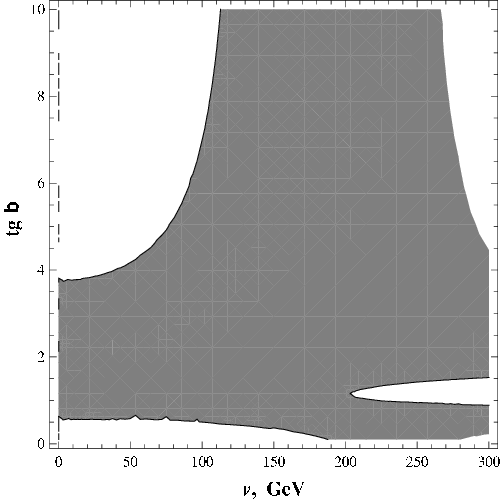}}
\end{picture}
\end{center}
\caption{ \small
Cross sections of the $\mu$-surface, see (\ref{angle}), at the temperature 
close to zero (right panel) and ${\tt tg}\beta=$1 (left panel, see also 
Fig.1).
White area in the right plot corresponds to the parameter values 
when the effective mass term (\ref{umass}) has a saddle configuration,
$A_{t,b}=$1800 GeV, $\mu=$2000 GeV. The $\beta$-surface is very close 
to the $\mu$-surface.
The squark sector parameter values are $m_Q=$500 GeV, $m_U=$200 GeV,
$m_D=$800 GeV, the charged Higgs boson mass $m_{H^\pm}=$150 GeV.
}
\label{projections}
\end{figure}
The presence of nonzero effective parameters $\lambda_6$ and $\lambda_7$ 
is essential to get the critical temperature of the order of 100 GeV.
Useful analytical approximation can be obtained
using the minimization conditions (\ref{diagmu1})-(\ref{diagmu2}), then the critical angle 
$\bar \beta (T)$ defined by (\ref{angle}) can be
expressed as
\footnote{In different context analogous relation between 
${\tt tg}\beta$ and ${\tt tg} \bar \beta$ can be 
found in \cite{brignole}, where the mass term of the form $v^2 
f(\beta,T)$ with $f(\beta,T)=a(\beta) T^2 - b(\beta)$ was analyzed  
for a special case of degenerate squark masses, $A=\mu=0$ and within the 
high-temperature expansion. 
Our quartic potential is very different from the tree-level 
$(g^2_1+g^2_2)/8 \, \lambda_T v^4 \cos^2 2\beta$ plus a logarithmic term 
\cite{brignole}, so the expression $3E/\lambda_T>$1 of 
Bochkarev-Shaposhnikov criteria $v(T_c)/T_c>$1 for the absence of 
sphaleron in the broken phase gives different results with nonzero 
threshold corrections.}
\begin{eqnarray}
{\tt tg} 2 \bar \beta & = & {\tt tg} 2 \beta \;
\frac{1}
{(\frac{v^2}{2m^2_A}-\alpha_1)} \;
\frac{1}
{\frac{2\lambda_1 \cos^2 \beta-2\lambda_2 \sin^2 \beta}{\cos 2\beta}
- \lambda_{345} + \frac{2m^2_A}{v^2} + \alpha_2}
\label{eqangle67}
\end{eqnarray}
where
\begin{equation}
\alpha_1=\frac{\lambda_5}{2}
           +\frac{1}{4}(\lambda_6 {\tt ctg} \beta+\lambda_7 {\tt tg}
\beta), \quad
\alpha_2= \lambda_6 ({\tt tg} 2\beta - {\tt ctg} \beta)
           -\lambda_7 ({\tt tg} \beta + {\tt tg} 2\beta).
\end{equation}
The assumption $\bar \beta(T) = \beta(0)$, i.e. only the modulo of $v(T)$
but not the direction in $(v_1, v_2)$ plane are changed in the process of
thermal evolution, gives for the critical angle (\ref{eqangle67})
\begin{eqnarray}
&-\frac{m^2_A}{v^2}(2\lambda_5+\lambda_6 \, {\tt ctg}\beta+\lambda_7 \,
{\tt tg}\beta)
+\frac{v^2}{m^2_A}
\; [ \frac{2\lambda_1 - 2\lambda_2 {\tt tg}^2 \beta
        + \lambda_6 (3 {\tt tg}\beta - {\tt ctg}\beta)
        + \lambda_7 ({\tt tg}^3 \beta -3 {\tt tg} \beta) }{1-{\tt tg}^2
\beta}
        - \lambda_{345} ] = 0.&
\label{tangent67}
\end{eqnarray}
This approximation may be too rough at small $m_{H^\pm}$, as pointed out 
in \cite{brignole}. The saddle configuration changes not only the 
shape, but also the horizontal orientation in the process of thermal 
evolution. For the case 
$\lambda_5=\lambda_6=\lambda_7=0$ (\ref{tangent67}) is reduced to
\begin{equation}
{\tt tg}^2 \beta = \frac{2\lambda_1 - \lambda_{345}}{2\lambda_2 -
\lambda_{345}}
\label{tangent}
\end{equation}
Combining (\ref{diagmu2}) and (\ref{tangent67}), where
only the leading power terms in $\lambda_i$ are kept and omitting
$\lambda_5 << m^2_A/v^2$,
the equation (\ref{eqT}) can be written in the form
\begin{equation}
\lambda_1 \, (2\lambda_2 - \lambda_{345})^2 \, 
+ \lambda_2 \, (2\lambda_1 - \lambda_{345})^2 \,
+  \lambda_{345} (2\lambda_1 - \lambda_{345}) (2\lambda_2 -
\lambda_{345})= 0
\label{lambdaeqT67}
\end{equation}
The vacuum expectation value $v$ and mass $m_A$ do not explicitly
participate in this equation, only the dimensionless effective parameters 
$\lambda_i$ of the quartic potential terms.
The left-hand side of (\ref{lambdaeqT67}) approaches zero from below as 
$v$ increases, demonstrating however no solution for the saddle configuration. 
This can be understood qualitatively if we rewrite (\ref{lambdaeqT67}) in 
the form $\lambda_1 {\tt ctg}^2\beta+\lambda_2 {\tt tg}^2\beta+
\lambda_{345}=$0 where the numerical values in 
the $\lambda$-pattern, see Fig.\ref{histlambda}, calculated in the BGX
scenario $\lambda_1<0$, $\lambda_2>0$ and $\lambda_{345}<0$, so in 
(\ref{tangent}) ${\tt tg} \beta < 1$.

Turning back to the case of effective field theory when the squarks decouple at the
$m_{top}$ energy scale, the evaluation of thermal masses of Higgs bosons, mixing angles and couplings can be done using results of \cite{physrev}. For example, the 
thermal evolution of the CP-even Higgs bosons $h$ and $H$ is expressed by
(compact notations $s_\alpha=\sin \alpha$, $c_\beta=\cos \beta$, etc. are 
used)
\begin{eqnarray}
m^2_h&=&c^2_{\alpha-\beta} m^2_A  
   + v^2 ( 2\lambda_1 s^2_{\alpha}
c^2_{\beta} + 2\lambda_2
c^2_{\alpha} s^2_{\beta}- 2( \lambda_3+ \lambda_4)c_{\alpha}
c_{\beta} s_{\alpha} s_{\beta}+ {\tt Re}\lambda_5
(s^2_{\alpha} s^2_{\beta} +c^2_{\alpha} c^2_{\beta})  \\
\nonumber&& - 2 c_{\alpha+\beta} ({\tt Re}\lambda_6
s_{\alpha} c_{\beta}
                    -{\tt Re}\lambda_7 c_{\alpha} s_{\beta})),\\
m^2_H&=&s^2_{\alpha-\beta} m^2_A   
 + v^2 ( 2\lambda_1 c^2_{\alpha} c^2_{\beta} +
  2\lambda_2
s^2_{\alpha} s^2_{\beta}+ 2(\lambda_3+
 \lambda_4)c_{\alpha}
c_{\beta} s_{\alpha} s_{\beta}+ {\tt Re}\lambda_5
(c^2_{\alpha} s^2_{\beta} +s^2_{\alpha} c^2_{\beta})  \\
\nonumber && +2 s_{\alpha+\beta} ({\tt Re} \lambda_6
c_{\alpha} c_{\beta}
                    +{\tt Re}\lambda_7 s_{\alpha} s_{\beta})) , 
\end{eqnarray}
where the mixing angle $\alpha$ of the CP-even states $h$ and $H$
is
\begin{eqnarray}
{\tt tg} 2\alpha &\hspace{1mm} =& \hspace{1mm}
\frac { s_{2\beta} m^2_A   - v^2 ((
\lambda_3 + \lambda_4) s_{2\beta}+2c^2_{\beta}
{\tt Re} \lambda_6
+ 2s^2_{\beta}  {\tt Re}\lambda_7) }
                        {c_{2\beta} m^2_A -v^2 (
2\lambda_1 c^2_{\beta} - 2\lambda_2 s^2_{\beta} - {\tt Re}
\lambda_5 c_{2\beta}+({\tt Re}\lambda_6- {\tt Re}
\lambda_7)s_{2\beta} )}\,.
\label{t2b}
\end{eqnarray}

\unitlength 1.00cm
\begin{figure}[h!]
\begin{center}
\begin{picture}(8,6)
\put(-4.,-0.5){\epsfxsize=7cm \epsfysize=7cm \leavevmode
\epsfbox{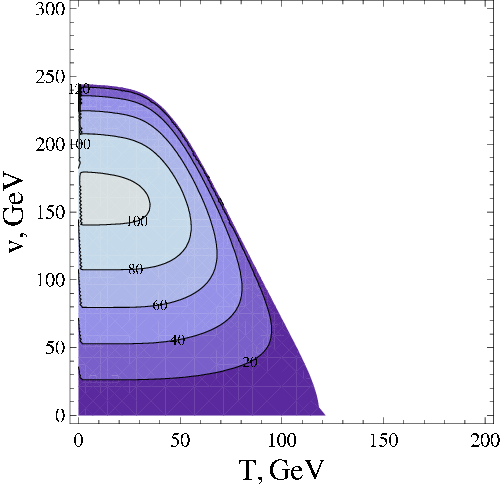}}
\put(4.5,-0.5){\epsfxsize=7cm \epsfysize=7cm \leavevmode
\epsfbox{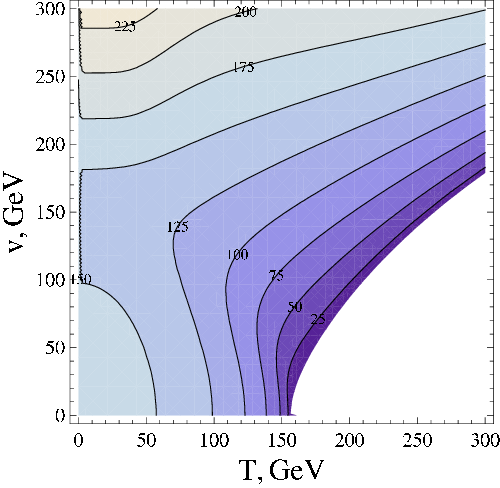}}
\end{picture}
\end{center}
\caption{ \small
In the shaded areas $m_h$ (left panel) and $m_H$ (right panel) are 
positively defined at the parameter values ${\tt tg} 
\beta=$5, $m_{H^\pm}=$180 GeV, $A_{t,b}=$1200 GeV, $\mu=$500 
GeV. Isocontours of constant $m_h$ and $m_H$ masses are indicated.
The squark sector parameter values are $m_Q=$500 GeV, $m_U=$200 GeV,
$m_D=$800 GeV.
}
\label{vthiggstgb5}
\end{figure}

\unitlength 1.00cm
\begin{figure}[h!]
\begin{center}
\begin{picture}(8,6)
\put(-4.,-0.5){\epsfxsize=7cm \epsfysize=7cm \leavevmode
\epsfbox{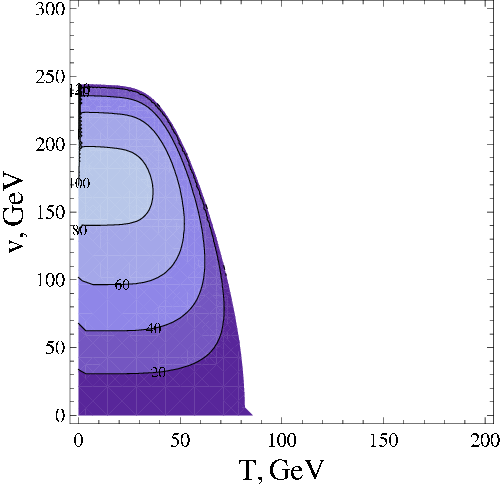}}
\put(4.5,-0.5){\epsfxsize=7cm \epsfysize=7cm \leavevmode
\epsfbox{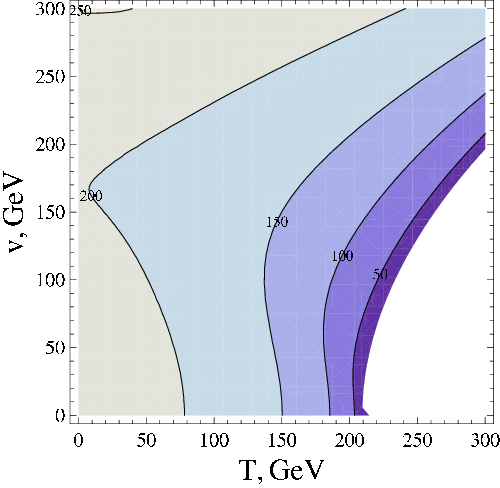}}
\end{picture}
\end{center}
\caption{\small
The same contours as in Fig.\ref{vthiggstgb5} at ${\tt tg}
\beta=$15, $m_{H^\pm}=$230 GeV
}
\label{vthiggstgb15}
\end{figure}

\unitlength 1.00cm
\begin{figure}[h!]
\begin{center}
\begin{picture}(8,6)
\put(-4.,-0.5){\epsfxsize=7cm \epsfysize=7cm \leavevmode
\epsfbox{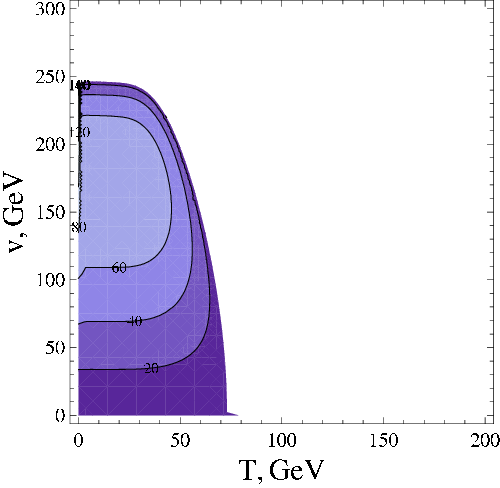}}
\put(4.5,-0.5){\epsfxsize=7cm \epsfysize=7cm \leavevmode
\epsfbox{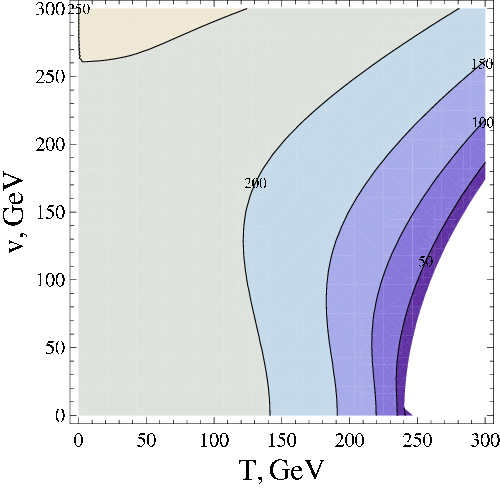}}
\end{picture}
\end{center}
\caption{\small
The same contours as in Fig.\ref{vthiggstgb5} at ${\tt tg}
\beta=$40, $m_{H^\pm}=$260 GeV
}
\label{vthiggstgb40}
\end{figure}

We show the regions of the 
($v$,$T$) plane where the CP-even Higgs boson masses $m_h$ and $m_H$ are 
positively defined in Fig.\ref{vthiggstgb5} - \ref{vthiggstgb40} 
(shaded areas). 
"Tachyonic" areas (shown in white colour) correspond to the negative 
squared 
masses of $m_h$ or $m_H$, see (\ref{masseig}), when the fluctuations
of physical fields $h$ and $H$ near the unstable local extremum ($v_1(T)$, 
$v_2(T))$ grow exponentially with time. Configuration of the 
effective potential $U_{eff} (h,H,A,T)$ expressed in the physical
fields $h$, $H$ in these areas is a saddle or a function with 
negative or indefinite sign values 
unbounded from below. The 'saddle' temperature $T_c=$120 GeV (see Fig.\ref{Ueff67zero}) is
close to the temperature, when the thermal mass $m_h(T)$ vanishes, only at the low ${{\tt tg}\beta}$
values. The heavy scalar mass $m_H(T)$ vanishes at the temperatures substantially higher than $T_c$.
In the scenario under consideration at higher ${{\tt tg}\beta}$ one should increase the charged scalar mass to respect the zero-temperature condition $v(0)=$246 GeV.

\section{Summary}

\noindent
Our analysis of the effective MSSM finite-temperature potential is based 
on a calculation of various one-loop temperature corrections from 
the squark-Higgs boson sector for the case of nonzero trilinear parameters
$A_t$, $A_b$ and Higgs superfield parameter $\mu$. 
Quantum corrections are incorporated in the parameters $\lambda_{1,...7}(T)$ 
of the effective two-doublet potential (\ref{eq:genU}), which is then 
explicitly rewritten in 
terms of Higgs boson mass eigenstates, using the approach developed
in \cite{physrev,echaya}. 
The effective parameters
$\lambda_1(T)$,...$\lambda_7(T)$ include the threshold corrections from
triangle, box and "fish" diagrams together with the logarithmic and the
wave-function renormalization terms. Dominant contribution comes from
the triangle and box graphs (Fig.\ref{thresh}, left and central) and can 
be written in a compact form by means of the generalized Hurwitz 
zeta-function. 

Temperature evolution of the potential $U_{eff}(v_1,v_2)$ expressed in 
terms of the background fields ($v_1(T)$,$v_2(T)$) is very sensitive to the 
MSSM scenario under consideration. We are using the scenarios with large
$A_{t,b}$ and $\mu$ (about/of the order of 1 TeV), favored by the available 
experimental data. Two characteristic sets, (A) and (B) (see section 3), are used for the squark-Higgs boson
sector parameters $m_Q$, $m_U$ and $m_D$. A relatively light stop quark is inherent
for the set (A), while with the set (B) sbottom quark is the lightest scalar quark eigenstate
(see Fig. \ref{mt1_mb1}) for an extensive regions of the MSSM parameter space.
Our analysis is essentially {\it two-dimensional}, i.e. the electroweak symmetry breaking
is considered for the two-dimensional potential surface $U(v_1 (T), v_2(T))$, the shape of which
is defined by nine parameters $\mu^1_1$, $\mu^2_2$, $\mu^2_{12}$ and $\lambda_1$,... $\lambda_7$ 
in the general (nonsupersymmetric) two-Higgs-doublet model. For the case of discrete Peccei-Quinn symmetry two
parameters are equal to zero: $\lambda_6=\lambda_7=$0. Using the terminology of the theory of
catastrophes \cite{catastrophe}, we analyse the local behavior of the two-dimensional potential function, dependent on the
several (less than five) control parameters. The surface of equilibrium points defined by the zero gradient
$\nabla U(v_1,v_2)=$0 looks like either a paraboloid function unbounded from above with the global minimum at the origin
or a saddle configuration. Two types of surfaces are separated by the critical condition 
${\tt det} \, \partial^2 U_{eff}(v_1,v_2)/ \partial v_i \partial v_j =$0 which defines the bifurcation set in the
MSSM parameter space. The electroweak phase transition along some direction in the $v_1$, $v_2$ plane occurs when the 
temperature evolution of $\lambda_{1,...5}$ from the TeV temperature scale down to zero temperature transforms a surface 
unbounded from above to a saddle configuration. 

For example, the surface of equilibrium points with the two 'flat directions' $v_1= \pm v_2$ for the tree-level MSSM 
potential at zero temperature is shown in Fig.\ref{susy_potential}.
When the non-temperature threshold corrections \cite{yaflast} to the 
tree-level zero temperature MSSM Higgs potential are included at some values of , $A_{t,b}$ and $\mu$, it 
changes the shape, becoming sign indefinite and unbounded from above, developing a saddle configuration. 

In section 3 four types of bifurcation sets for the two-Higgs-doublet potential $U_{eff}(v_1,v_2)$ are found.
The bifurcation set (1) defined by Eq.(\ref{uneq}) develops a phase transition in the
direction which is fixed by Eq.(\ref{angle1}) in the ($v_1,v_2$) plane. In the MSSM only the parameter set (A) of the squark-Higgs bosons sector,
characterized by the light stop quark, shows the necessary configuration of the surface for stationary points (paraboloid
with a global minimum at the origin at high temperatures and a saddle at low 
temperatures\footnote{The LHC exclusion contours \cite{lhc} for the gluino, squark and wino masees based on the luminosity 35 pb$^{-1}$ at $\sqrt{s}=$ 7 TeV in the GMSB and mSUGRA scenarios have yet no significant impact on the full MSSM parameter space, but reduce a parametric configurations to some extent.}). The bifurcation contour (also called the separatrix in the catastrophe theory terminology) in the ($A$, $\mu$) plane for the set (1) is shown in Fig.\ref{conditionlambdas}. The parameter set (B) with
the light sbottom always gives a saddle configuration because $\lambda_1 < $0 and $\lambda_2 > $0 in the parameter space. 
The bifurcation sets (2) and (3) are
similar, demonstrating a phase transition in the direction defined by Eq.(\ref{angle34}) in the ($v_1,v_2$) plane. The bifurcation contours 
in the ($A$, $\mu$) plane for the set (2) are shown in Fig.\ref{conditionlambdas_2}. Again, only the parameter set (A)
with the light stop demonstrates the necessary configuration of the equilibrium surfaces. The bifurcation set (4)
includes a phase transition in the direction of Eq.(\ref{angle}) at the temperature defined by Eq.(\ref{eqT}). This case
was analyzed earlier in the literature in the context of the one-dimensional effective potential. The bifurcation contours
for the case of parameter set (A) are shown in Fig. \ref{projections}. Summarizing, in all four cases the global minimum
at the origin $v_1=v_2=$0, $U_{eff} (0, 0) = $0 at high temperatures is transformed to a local minimum with $U_{eff} (v_1, v_2) < $0 at a lower temperature for the parameter set (A), but the directions of transition to this minimum in the ($v_1,v_2$) plane are different.

Oscillations of Higgs fields in the vicinity of an extremum ($v_1(T),v_2(T)$) give the effective potential 
with a minimum moving along the surface of stationary points. 
The potential $U_{eff}(h,H,A)$ written in terms of physical Higgs fields
(i.e. Higgs mass eigenstates $h$,$H$ and $A$) demonstrates the spectrum of scalars with positively 
defined masses which are reaching zero at different temperatures, 
see Figs.\ref{vthiggstgb5} - \ref{vthiggstgb40}.  These temperatures
are, as a rule, not close to the 'critical' temperature $T_c$, when
the potential $U_{eff}(v_1,v_2)$ forms a horizontal 'narrow gully',
so not only the first, but also the second derivatives are zero in some direction. 

The isocontours for CP-even scalar masses $m_h$ and $m_H$
fall down in an extremely narrow temperature region
near $T=$0, see Figs.\ref{vthiggstgb5} - \ref{vthiggstgb40}, so during
the 'overturn' of the potential a nearly step-like decrease of 
$v(T)$ must happen to keep constant masses.

Although our evaluation is performed at the one-loop only and not all 
possible corrections are considered, usually it is possible 
to adjust $\mu$, $A_{t,b}$ (or $X_{t,b}$), $m_Q$, $m_U$ and $m_D$ 
of the MSSM parameter space in such
a way that the boundary condition for zero temperature $v(0)=$246 GeV
is respected, the lightest Higgs boson mass is large enough, the critical temperature is of the
order of 100 GeV or higher, and the phase transition is of the first 
order. Threshold corrections with $\mu$, $A_{t,b}$ of the order of 1 TeV can increase the strength the electroweak phase transition.
Independently on the temperature evolution scenarios, one should not forget 
that the value of ${\tt tg}\beta$ at zero 
temperature must be consistent with the range from 5 to 
50-60, provided by phenomenological restrictions from LEP2 and
Tevatron data for the reactions $e^+ e^-
\to hZ$, $h\to b \bar b$, and $pp\to t\bar t$, $t\to H^\pm b$.
In the nearest future useful information about the allowed regions of the MSSM parameter space
could be provided by the LHC Higgs physics program \cite{higgslhc}.
Availability of the criteria $v(T_c)/T_c \sim$1 for the absence of sphaleron in the broken phase deserves a more careful
study with more precise evaluation of radiative corrections from other sources,
especially the infrared ones.

Only the case of real-valued MSSM parameters $A_{t,b}$ and 
$\mu$ was considered. Generalization to the complex-valued parameters (the case 
of explicit CP violation in the squark-Higgs and the two-doublet Higgs 
sectors) is straightforward with radiation corrections defined by 
Eqs.(\ref{eq:lambda5})-(\ref{eq:lambda7}), where phases of $A_{t,b}$ 
and $\mu$ can be introduced. Complex $\lambda_{5,6,7}$ lead to the mixing 
of CP-even $h$,$H$ and CP-odd $A$ scalars resulting in the Higgs bosons 
without a definite CP-parity $h_1$,$h_2$ and $h_3$ with specific 
properties, modifying the qualitative picture described above.

\vskip 0.5cm
\noindent
{\large \bf Acknowledgements}
\vskip 0.2cm
\noindent
M.D.(MSU) is grateful to Mikhail Shaposhnikov for useful discussion. 
Work was partially supported by grants ADTP 3341, RFBR 10-02-00525-a, NS 1456.2008.2
and FAP contract 5163.

\newpage
\section{Appendix}

Inputs and some details of various quantum corrections calculation, 
see Eq.(\ref{eq:lambda1}) and the following formulas, are given below.
Supersymmetric potential of the Higgs bosons - third generation of scalar 
quarks interaction has the form \cite{haberhempf}
\begin{equation} 
{\cal V}^{\,0} =
{\cal V}_M + {\cal V}_\Gamma + {\cal V}_\Lambda + {\cal
V}_{\widetilde Q}\,, \label{eq:HHpot} 
\end{equation}
where
\begin{equation} 
{\cal V}_M =
-\mu_{ij}^2\Phi_i^{\dag}\Phi_j+ M_{\widetilde
Q}^{\,2}\left(\widetilde
  Q^{\,\dag}\widetilde Q\right)
+M_{\widetilde U}^{\,2}\widetilde U^*\widetilde U +M_{\widetilde
D}^{\,2}\widetilde D^*\widetilde D\,, \end{equation}
\begin{equation} {\cal V}_\Gamma =
\Gamma_i^D\left(\Phi^{\dag}_i\widetilde Q\right)\widetilde D
+\Gamma_i^U\left(i\Phi_i^T\sigma_2\widetilde Q\right)\widetilde U
+\stackrel{*}{\Gamma_i^D}\left({\widetilde
Q}^{\,\dag}\Phi_i\right)\widetilde D^*
-\stackrel{*}{\Gamma_i^U}\left(i\widetilde
Q^{\,\dag}\sigma_2\Phi_i^*\right) \widetilde U^* \, ,
\label{eq:VGamma} \end{equation} \begin{equation} {\cal V}_\Lambda =
\Lambda_{ik}^{jl}\left(\Phi^{\dag}_i\Phi_j\right)
\left(\Phi^{\dag}_k\Phi_l\right) +\left(\Phi^{\dag}_i\Phi_j\right)
\left[\Lambda_{ij}^Q\left(\widetilde Q^{\,\dag}\widetilde Q\right)
+\Lambda_{ij}^U \widetilde U^*\widetilde U +\Lambda_{ij}^D
\widetilde D^*\widetilde D\,\right] + \label{eq:VL} \end{equation}
$$\quad+
\overline\Lambda_{\,ij}^{\,Q}\left(\Phi^{\dag}_i\widetilde Q\right)
\left(\widetilde Q^{\,\dag}\Phi_j\right)
+\frac{1}{2}\left[\Lambda\epsilon_{ij}
\left(i\Phi_i^T\sigma_2\Phi_j\right)\widetilde D^*\widetilde
U+{\mbox {h.c.}}\right]\,, \quad i,j,\,k,l=1,2 \,, $$
${\cal V}_{\widetilde Q}$ denotes the four scalar quarks 
interaction terms,
$\sigma_2\equiv\left(\begin{array}{cc} 0 & i\\
- i & 0\end{array}\right)$, and $\Lambda^I$ are defined by the
tree-level equalities
$$\Lambda^Q = {\tt diag}\{\frac{1}{4}(g^2_2-g_1^2 Y_Q), \quad
h_U^2-\frac{1}{4}(g^2_2-g_1^2 Y_Q)\},$$
$$\overline{\Lambda}^Q = {\tt diag}\{h_D^2-\frac{1}{2}g^2_2,
\frac{1}{2}g^2_2-h_U^2\}, \qquad \Lambda^U = {\tt
diag}\{-\frac{1}{4}g^2_1 Y_U, h_U^2+\frac{1}{4}g_1^2 Y_U\},$$
$$\Lambda^D = {\tt diag}\{h_D^2-\frac{1}{4}g_1^2 Y_D, \frac{1}{4}g_1^2
Y_D\},\qquad \Lambda = -h_U h_D.$$ 
Here the squark hypercharges are $Y_{Q_i}=1/3(-1)$, $Y_{D_i}=2/3(2)$, 
$Y_{U_i}=-4/3$, and
the Yukawa couplings for the third generation squarks 
$h_{\,t} = \frac{\sqrt{2}\, m_{\,t}}{v \sin\beta },\, h_{\,b} =
\frac{\sqrt{2}\, m_b }{v \cos\beta }\,.$ In the general case of
complex-valued parameters
\begin{equation} 
\Gamma_{\{1; \,2\}}^{\,U} = h_U\, \{-\mu^*; A_U\}, \qquad
\Gamma_{\{1; \,2\}}^{\,D} = h_D \,\{A_D\,; -\mu^*\} ,
\label{eq:Gammi} 
\end{equation}
In order to calculate, e.g., the one-loop threshold corrections
(\ref{eq:lambda1})-(\ref{eq:lambda7}) first we extract from the
potential (\ref{eq:HHpot}) the triple and quartic interactions presented 
in Fig.\ref{fg:vertix3} and Fig.\ref{fg:vertix4}. The triangle and 
box diagrams which 
contribute, for example, to the threshold corrections included in 
$\lambda_1$ are shown in Fig.\ref{fg:diagsL1} together with symbolic 
expressions for the temperature one-loop integrals. Their sum multiplied 
by the color factor 3 gives $\lambda^{\it thr}_1$, see (\ref{eq:lambda1}).
\unitlength 1mm
\begin{figure}
\begin{center}
\begin{picture}(40,140)
\put(-85,-85){\epsfxsize=21cm \epsfysize=27cm \leavevmode
\epsfbox{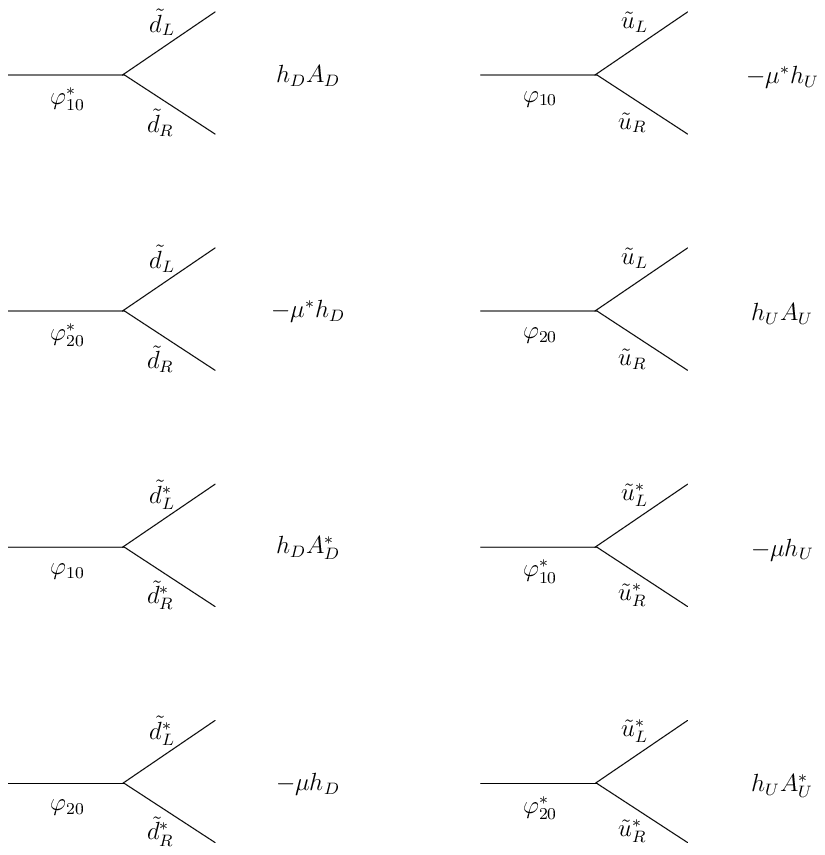}}
\end{picture}
\end{center}
\vskip 25mm
\caption{\small
The triple interactions extracted from the squark-Higgs sector, 
see (\ref{eq:HHpot}).
\label{fg:vertix3}}
\end{figure}

\begin{figure}
\begin{center}
\begin{picture}(40,140)
\put(-100,-70){\epsfxsize=21cm \epsfysize=27cm \leavevmode
\epsfbox{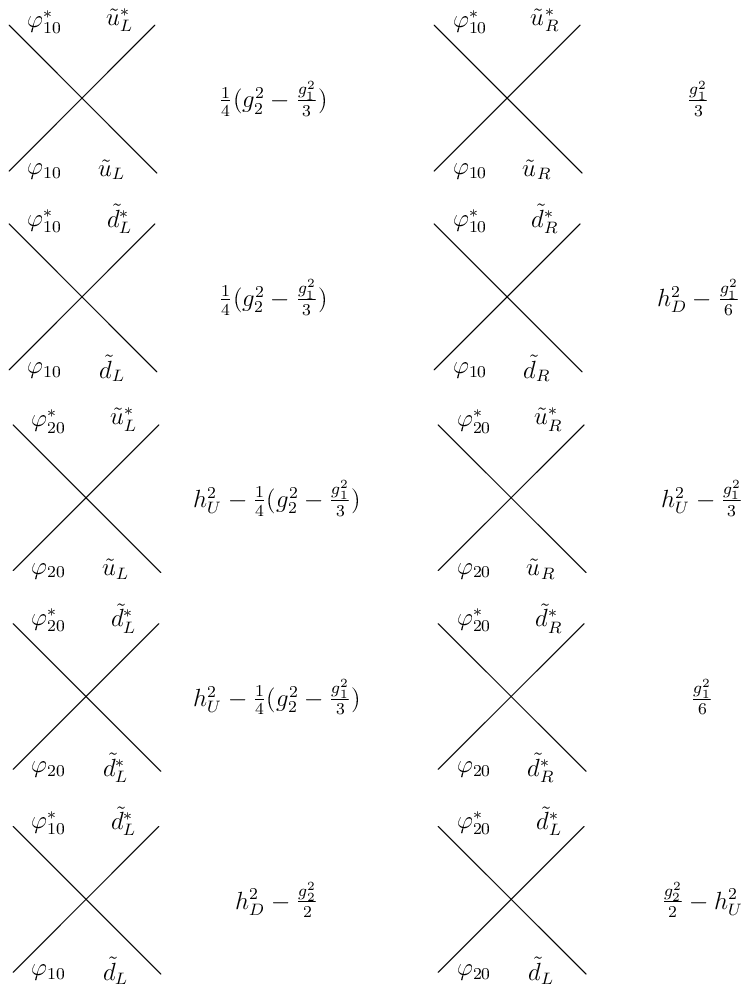}}
\end{picture}
\end{center}
\vskip 20mm
\caption{\small
The quartic interactions extracted from the squark-Higgs sector,
see (\ref{eq:HHpot}).
\label{fg:vertix4}}
\end{figure}

\begin{figure}
\begin{center}
\begin{picture}(40,190)
\put(-95,-20){\epsfxsize=19cm \epsfysize=25cm \leavevmode
\epsfbox{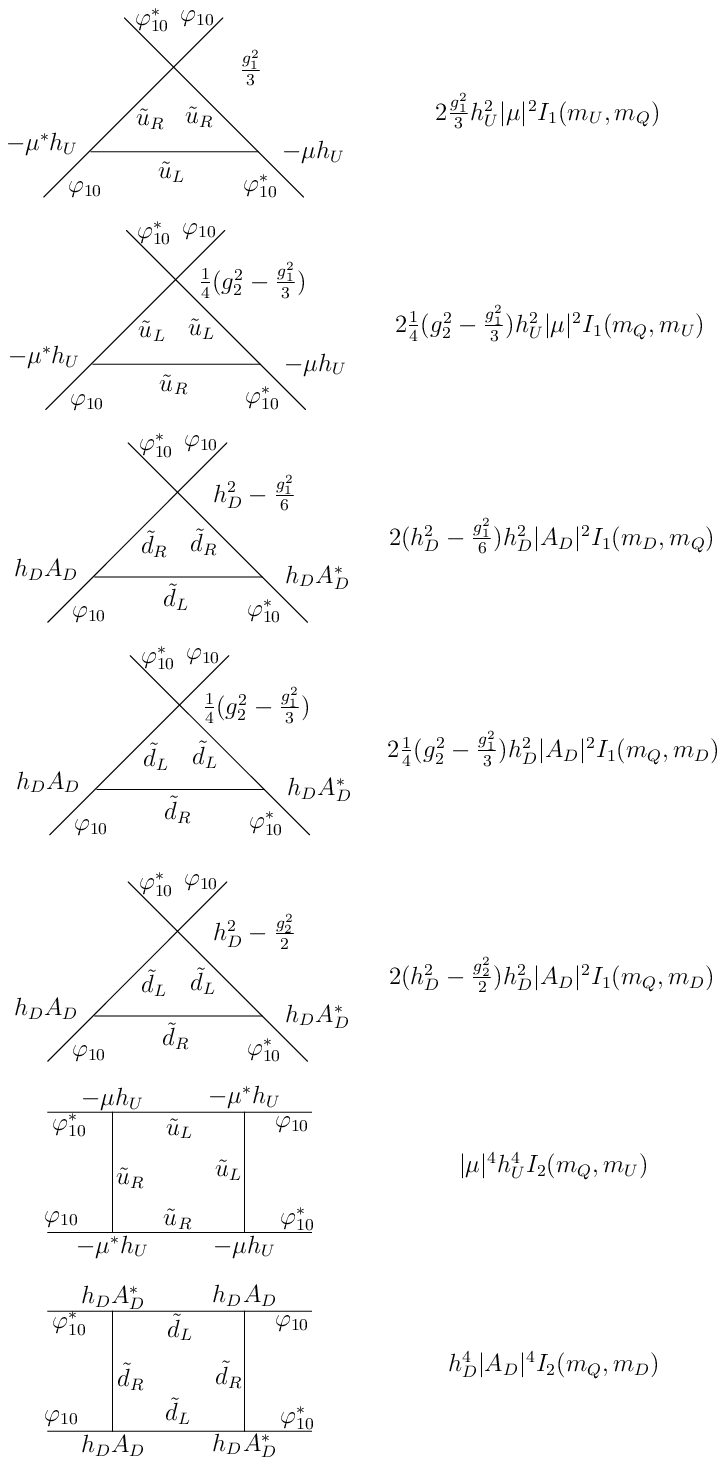}}
\end{picture}
\end{center}
\caption{\small
Triangle and box diagrams contributing to $\lambda^{thr}_1$, see
(\ref{eq:lambda1}).
\label{fg:diagsL1}}
\end{figure}

\newpage
\pagestyle{plain}
{


\newpage

\unitlength 1.00cm
\begin{figure}[t]
\linethickness{0.4pt}
\begin{center}
\begin{picture}(9,14)
\put(-2.5,1.0){\epsfxsize=14cm \epsfysize=12cm \leavevmode
\epsfbox{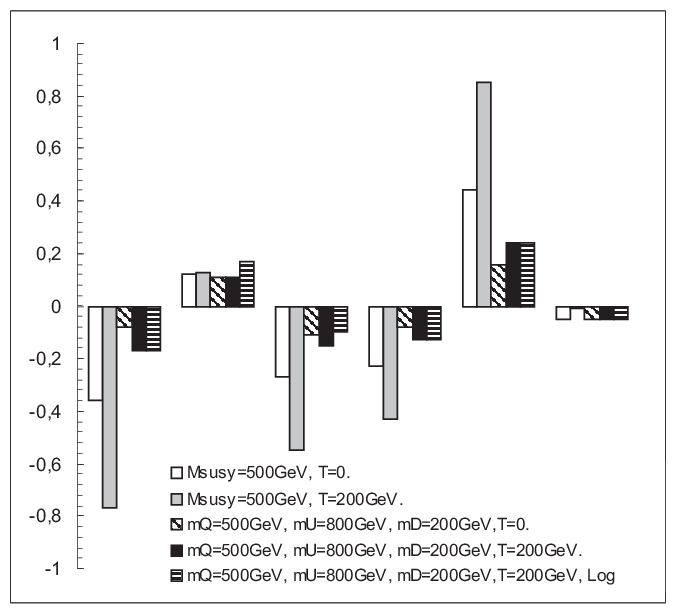}}
\end{picture}
\end{center}
\caption{\small
Histograms for temperature-dependent $\lambda_i$ ($i$=1,...7) with various 
quantum corrections in the framework of the CPX-like scenario \cite{cpx},
$A_t=A_b=$1000 GeV, $\mu=$2000 GeV, in the cases of (a) degenerate squark 
masses $m_Q=m_t=m_b=M_{SUSY}=$500 GeV, zero temperature, (b) degenerate 
squark masses $m_Q=m_t=m_b=M_{SUSY}=$500 GeV, $T=$200 GeV, (c) different 
squark masses $m_Q=500$ GeV,Ð$m_t =800$ GeV,; $m_b = 200$ GeV at zero 
temperature, and (d) different squark masses $m_Q=500$ GeV, $m_t =800$ 
GeV, $m_b = 200$ GeV at $T=$200 GeV.
}
\label{histlambda}
\end{figure}

\unitlength 1.00cm
\begin{figure}[t]
\linethickness{0.4pt}
\begin{center}
\begin{picture}(10,10)
\put(-2.5,1.0){\epsfxsize=14cm \epsfysize=12cm \leavevmode
\epsfbox{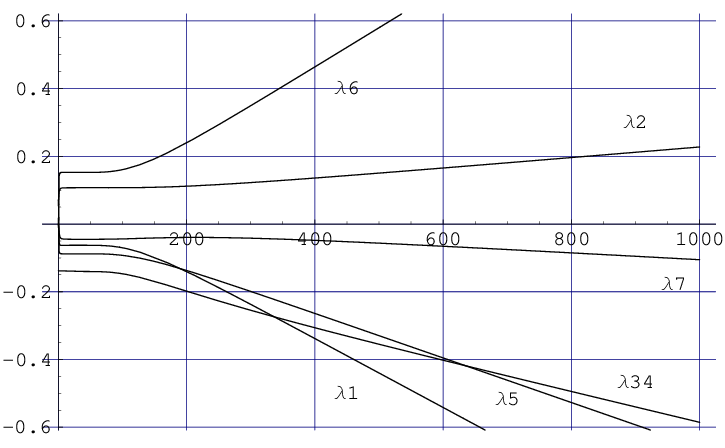}}
\end{picture}
\end{center}
\caption{\small  
Effective temperature-dependent parameters $\lambda_i$ ($i$=1,...7)
with the one-loop threshold and logarithmic corrections at
$m_Z =91.19$ GeV, $m_b=3$ GeV, $m_t=175$ GeV, $m_W=79.96$ GeV, 
$g_2=0.6517$, $g_1=0.3573$, $G_F=1.174\cdot10^{-5}$ GeV$^{-2}$,
$M_{SUSY}=500$ GeV, $m_Q=500$ GeV, $m_t = 800$ GeV, $m_b = 200$ GeV,
$\mu=$2000 GeV, $A=X_t+\mu/\tan\beta$, $X_t=700$ GeV, $\tan\beta=5$,
$h_t=1$, $h_b=0.1$.
}
\label{lambda}
\end{figure}


\normalsize

\begin{thebibliography}{99}

\bibitem{ewtrans_gen}
for a review, see: V.A. Rubakov, M.E. Shaposhnikov, Usp.Fiz.Nauk {\bf 166}
(1996) 493 (hep-ph/9603208)\\
A.G. Cohen, D.B. Kaplan, A.E. Nelson,
Ann.Rev.Nucl.Part.Sci. {\bf 43} (1993) 27 (hep-ph/9302210)\\
A.D. Dolgov, Phys.Rep. {\bf 222} (1992) 309

\bibitem{kirzhnits}
A. Linde, Rep.Prog.Phys. {\bf 42} (1979) 389\\
D. Kirzhnits, A. Linde, Ann.Phys. {\bf 101} (1976) 195\\
S. Weinberg, Phys.Rev. D{\bf 9} (1974) 3357\\ 
L. Dolan, R. Jackiw, Phys.Rev. D{\bf 9} (1974) 3320

\bibitem{approaches}
L. Fromme, S. Huber, M. Seniuch, JHEP 0611 (2006) 038 (hep-ph/0605242)\\ 
Y. Okada et al., in: Proc. of CERN Workshop on
CP studies and nonstandard Higgs physics, ed. by S.Kraml, G.Azuelos,
D.Dominici, J.Ellis, G.Grenier, H.Haber, J.S.Lee, D.Miller, A.Pilaftsis,
W.Porod, CERN Yellow Report 2006--009, 2006 (hep-ph/0608079)\\
K. Funakubo, S. Tao, F. Toyoda, Prog.Theor.Phys. {\bf 109} (2003) 415 
(hep-ph/0211238)\\
M. Losada, Nucl. Phys. B{\bf 569} (2000) 125 (hep-ph/9905441)\\
M. Laine, K. Rummukainen, Nucl.Phys. B{\bf 545} (1999) 141
(hep-ph/9811369)\\
M. Brhlik, G.J. Good, G.L. Kane, Phys.Rev. D{\bf 63} (2001) 035002
(hep-ph/9911243) \\ 
J. Cline, K. Kainulainen, Nucl.Phys. B{\bf 510} (1998) 88
(hep-ph/9705201)\\
P. Arnold, O. Espinosa,
Phys.Rev. D{\bf 47} (1993) 3546 (Erratum-ibid. D50 (1994) 6662)
(hep-ph/9212235) \\
G.F. Giudice, Phys.Rev. D{\bf 45} (1992) 3177\\
N. Turok, J. Zadrozny, Nucl. Phys. B{\bf 369} (1992) 729\\
A.I. Bochkarev, S.V. Kuzmin, M.E. Shaposhnikov, Phys.Lett. B{\bf 244} 
(1990) 275

\bibitem{window}
M. Carena, G. Nardini, M. Quiros, C.E.M. Wagner. Nucl.Phys. {\bf B812} (2009) 243 (arXiv:0809.3760 [hep-ph])\\
M. Carena, G. Nardini, M. Quiros and C.E.M. Wagner, JHEP 10 (2008) 062 (arXiv:0806.4297 [hep-ph])\\
M. Carena, M. Quiros, C.E.M. Wagner, Phys.Lett. B{\bf 380} (1996) 81 (hep-ph/9603420)

\bibitem{first_criteria}
A.I. Bochkarev, M.E.Shaposhnikov, Mod.Phys.Lett. A{\bf 2} (1987) 417\\
M.E. Shaposhnikov, JETP Lett. {\bf 44} (1986) 465

\bibitem{brignole}
A. Brignole, J.R. Espinosa, M. Quiros, F. Zwirner,
Phys. Lett. B{\bf 324} (1994) 181 (hep-ph/9312296)

\bibitem{montecarlo}
K. Kajantie, M. Laine, K. Rummukainen, M. Shaposhnikov,
Nucl.Phys. B{\bf 466} (1996) 189 (hep-lat/9510020)\\
K. Kajantie, K. Rummukainen, M. Shaposhnikov,
Nucl.Phys. B{\bf 407} (1993) 356 (hep-ph/9305345)

\bibitem{irproblem}
D. Gross, R. Pisarski, L. Yaffe, Rev.Mod.Phys. {\bf 53} (1981) 43\\
A. Linde, Phys.Lett. B{\bf 96} (1980) 289

\bibitem{KajLaiRumSha1995}
K. Kajantie, M. Laine, K. Rummukainen, M. Shaposhnikov,
Nucl.Phys. B{\bf 458} (1996) 90 (hep-ph/9508379)\\
K. Farakos, K. Kajantie, K. Rummukainen, M. Shaposhnikov,
Nucl.Phys. B{\bf 425} (1994) 67 (hep-ph/9404201)

\bibitem{dimred}
A. Jakovac, K. Kajantie, A. Patkos, Phys.Rev. D{\bf 49} (1994) 6810\\
S. Nadkarni, Phys.Rev. D{\bf 27} (1983) 917\\
T. Appelquist, R.Pisarski,  Phys.Rev. D{\bf 23} (1981) 2305\\
P. Ginsparg, Nucl.Phys. B{\bf 170} (1980) 388

\bibitem{laine}
M. Laine, Nucl.Phys. B{\bf 481} (1996) 43 (hep-ph/9605283)

\bibitem{losada}
G.R. Farrar, M. Losada, Phys.Lett. B{\bf 406} (1997) 60 (hep-ph/9612346)\\
M. Losada, Phys.Rev. D{\bf 56} (1997) 2893 (hep-ph/9605266)

\bibitem{cline}
J. Cline, K. Kainulainen, Nucl.Phys.B{\bf 482} (1996) 73 (hep-ph/9605235)

\bibitem{physrev}
E. Akhmetzyanova, M. Dolgopolov, M. Dubinin, Phys.Rev. D{\bf 71} (2005) 
075008 (hep-ph/0405264)

\bibitem{echaya}
E. Akhmetzyanova, M. Dolgopolov, M. Dubinin, Phys.Part.Nucl. {\bf 37} (2006) 677

\bibitem{other}
J.F. Gunion, H.E.Haber, Phys.Rev. D{\bf 67} (2003) 075019 
(hep-ph/0207010)\\
M. Dubinin, A. Semenov, Eur.Phys.J. C{\bf 28} (2003) 223 
(hep-ph/0206205)\\
F. Boudjema, A. Semenov, Phys.Rev. D{\bf 66} (2002) 095007 
(hep-ph/0201219)

\bibitem{LV97} 
L. Vergara, Journal of Physics  A{\bf 30} (1997) 6977

\bibitem{Amore2005} 
P.Amore, Journal of Physics A{\bf 38} (2005) 6463 (hep-th/0503142)

\bibitem{hurwitz}
P. Amore, Journal of Mathematical Analysis and
Applications {\bf 323} (2006) 63\\
M. Abramowitz and I. Stegun, Handbook of Mathematical Functions,
Dover Publications, NY, 1964

\bibitem{haberhempf}
H. Haber, R. Hempfling, Phys.Rev. D{\bf 48} (1993) 4280 (hep-ph/9307201)

\bibitem{choidreeslee}
S.Y. Choi, M. Drees, J.S. Lee, Phys.Lett. B{\bf 481} (2000) 57 
(hep-ph/0002287)

\bibitem{collins}
J.C. Collins, {\it Renormalization}, Cambridge Univ. Press, Cambridge,
1984

\bibitem{yaflast}
E. Akhmetzyanova, M.Dolgopolov, M. Dubinin,
Phys.Atom.Nucl. {\bf 70} (2007) 1549 (Yad.Fiz. {\bf 70} (2007) 1594)

\bibitem{bgx}
C. Balazs, M. Carena, A. Menon, D.Morrissey, C.E.M. Wagner, Phys.Rev. 
D{\bf 71} (2005) 075002 (hep-ph/0412264)\\
S.Heinemeyer, M.Velasco, in: Proc.of 2005 ILC Workshop, Stanford, USA, 
hep-ph/0506267

\bibitem{cpx}
M.~Carena, J.~Ellis, A.~Pilaftsis, C.~Wagner, Phys.Lett. B{\bf 495}
(2000) 155 (hep-ph/0009212)

\bibitem{catastrophe}
R. Gilmore, {\it Catastrophe theory for scientists and engineers}, John Wiley \& Sons, New York-Chichester-Brisbane-Toronto, 1981\\ 
V.I. Arnold, {\it Critical points of smooth functions and their canonical forms}, Uspekhi Math. Nauk (USSR), {\bf 30} (1975) 3\\
R. Thom, {\it Structural stability and morphogenesis}, Reading, Benjamin, 1975\\
M. Morse, {\it The critical points of a function of n variables}, Trans. Am. Math. Soc., {\bf 33} (1931) 72

\bibitem{PQsymm}
R. Peccei, H. Quinn, Phys.Rev.Lett. {\bf 38} (1977) 1440

\bibitem{kusenko}
A. Kusenko, P. Langacker, G. Segre, Phys.Rev. D{\bf 54} (1996) 5824 
(hep-ph/9602414)

\bibitem{affleck}
I. Affleck, M. Dine, Nucl.Phys. B{\bf 249} (1985) 361

\bibitem{wolfram}
S. Wolfram, {\it Mathematica} symbolic manipulation package, see~{\tt
http:{\protect //}www.wolfram.com}

\bibitem{lhc}
S. Chatrchyan et al. (CMS Collaboration), arXiv:1105.3152[hep-ex]; arXiv:1104.3168[hep-ex];
arXiv:1103.0953[hep-ex]; Phys.Lett. B698 (2011) 196 (arXiv:1101.1628[hep-ex]); Phys.Rev.Lett. 106 (2011) 011801 (arXiv:1011.5861[hep-ex]) 

\bibitem{higgslhc}
S. Abdullin et al.,  Eur.Phys.J.C{\bf 39S2} (2005) 41

\end{thebibliography}
\end{document}